\documentclass[english,onecolumn, draftcls]{IEEEtran}
\usepackage[T1]{fontenc}
\usepackage{verbatim}
\usepackage{units}
\usepackage{amsthm}
\usepackage{amsmath}
\usepackage{amssymb}
\usepackage{graphicx}

\makeatletter
 \let\oldforeign@language\foreign@language
 \DeclareRobustCommand{\foreign@language}[1]{%
   \lowercase{\oldforeign@language{#1}}}
\theoremstyle{plain}
\newtheorem{thm}{\protect\theoremname}
\theoremstyle{remark}
\newtheorem{rem}[thm]{\protect\remarkname}
\theoremstyle{plain}
\newtheorem{cor}[thm]{\protect\corollaryname}
\theoremstyle{definition}
\newtheorem{defn}[thm]{\protect\definitionname}
\theoremstyle{plain}
\newtheorem{lem}[thm]{\protect\lemmaname}
\theoremstyle{plain}
\newtheorem{prop}[thm]{\protect\propositionname}

\newtheorem{property}{Property}

\@ifundefined{showcaptionsetup}{}{%
 \PassOptionsToPackage{caption=false}{subfig}}
\usepackage{subfig}
\makeatother

\usepackage{babel}
\providecommand{\corollaryname}{Corollary}
\providecommand{\definitionname}{Definition}
\providecommand{\lemmaname}{Lemma}
\providecommand{\propositionname}{Proposition}
\providecommand{\remarkname}{Remark}
\providecommand{\theoremname}{Theorem}

\begin{document}

\title{The Wideband Slope of Interference Channels: The Large Bandwidth
Case}

\author{Minqi Shen, Anders H{\o}st-Madsen%
\thanks{The authors are with the Department of Electrical Engineering, University
of Hawaii Manoa, Honolulu, HI 96822 (e-mail: \{minqi,ahm\}@hawaii.edu.
This work was supported in part by NSF grants CCF 0729152 and CCF
1017823. This paper was presented in part at the 48th annual Allerton
Conference on Communication, Control and Computing, October 2010 (Urbana-Champaign,
IL).%
}}

\maketitle
\global\long\def\snr{\mathrm{SNR}}
\global\long\def\kuser{K-\mathrm{user}}
\global\long\def\trace{\mathrm{Tr}}
\global\long\def\var{\mathrm{var}}
\global\long\def\cov{\mathrm{cov}}
\global\long\def\ebno{\frac{E_{b}}{N_{0}}}
\global\long\def\slope{\mathcal{S}}
\global\long\def\mod{\mathrm{mod}}
\global\long\def\sinc{\mathrm{sinc}}

\global\long\def\ebnomin{\left.\frac{E_{b}}{N_{0}}\right|_{\min}}

\begin{abstract}
It is well known that minimum received energy per bit $\ebnomin$
in the interference channel is $-1.59dB$ as if there were no interference.
Thus, the best way to mitigate interference is to operate the interference
channel in the low-$\mathrm{SNR}$ regime. However, when the SNR is
small but non-zero, $\ebnomin$ alone does not characterize performance.
Verdu introduced the wideband slope $\mathcal{S}_{0}$ to characterize
the performance in this regime. We show that a wideband slope of $\frac{\mathcal{S}_{0}}{\mathcal{S}_{0,\mbox{no interference}}}=\frac{1}{2}$
is achievable. This result is similar to recent results on degrees
of freedom in the high SNR regime, and we use a type of interference
alignment using delays to obtain the result. We also show that in
many cases the wideband slope is upper bounded by $\frac{\mathcal{S}_{0}}{\mathcal{S}_{0,\mbox{no interference}}}\leq\frac{1}{2}$
for large number of users $K$.\end{abstract}
\begin{IEEEkeywords}
Interference channels, wideband slope, interference alignment.
\end{IEEEkeywords}

\markboth{IEEE Transactions on Information Theory}{M. Shen and A. H{\o}st-Madsen.}

\section{Introduction}

\PARstart{R}{ecently} there has been much interest in interference
channels \cite{AnnVee08,MotKha08IT,ShaKraChe07IT}. In \cite{CadambeJafar07}
it was shown that in the high-SNR regime, it is possible to achieve
$K/2$ degrees of freedom in a $K$-user interference channel (half
of the $K$ degrees of freedom if there were no interference). The
basic idea is to align interference from all $K-1$ undesired users
in half the signal space, and then receive the desired signal in the
other half space without interference, an idea pioneered by \cite{MaddahAliAl06}.
The paper \cite{CadambeJafar07} has inspired a large body of research
on interference alignment in the high-SNR regime, for example \cite{GomCadJaf08,CadJafSha09IT,CadambeJafarAl09,MotGhaKha09CoRR,EtkOrd09IT,GhaMotKha09CoRR}.

In this paper we consider the interference channel in the low-SNR
regime, where explicitly
\begin{eqnarray}
\snr & \triangleq & \frac{P}{BN_{0}},\label{eq:def snr}
\end{eqnarray}
 $P$ is the input power, and $B$ is the system bandwidth. While
the work in \cite{CadambeJafar07} and follow-up work shows impressively
that much can be done to mitigate the effect of interference in the
high-SNR regime, one could argue that the best way to mitigate the
effect of interference is to avoid the high-SNR regime and instead
operate in the low-SNR regime, when possible. It is well-known (e.g.,
\cite{Verdu90}) that in a point-to-point channel the received minimum
energy per bit $\left.\ebno\right|_{\min}=-1.59dB$ is achieved as
the spectral efficiency (bits/s/Hz) $R\to0$. It is also known from
\cite{Verdu90} that this energy is unchanged in the presence of interference.
Thus, in this limit the effect of interference is completely eliminated.
However, as Verdu pointed out in \cite{Ver02IT}, in practical systems
the spectral efficiency must be non-zero, though it might still be
small. One way to characterize the effect of this is through the \emph{wideband
slope}. The wideband slope is defined by
\begin{eqnarray}
\mathcal{S}_{0} & \triangleq & \lim_{\ebno\downarrow\ebno_{\min}}\frac{R\left(\ebno\right)}{10\log_{10}\ebno-10\log_{10}\ebnomin}10\log_{10}2,\label{eq:slope}
\end{eqnarray}
where $R\left(\ebno\right)$ is the spectral efficiency as a function
of $\ebno$. The wideband slope essentially represents a second order
approximation in the low power regime of the spectral efficiency as
a function of SNR, or first order approximation of the spectral efficiency
as a function of $\ebno$. For example, we can write
\begin{eqnarray*}
R & \approx & \frac{\mathcal{S}_{0}}{10\log_{10}2}\left(10\log_{10}\ebno-10\log_{10}\ebnomin\right)\\
10\log_{10}\ebno & \approx & 10\log_{10}\ebnomin+\frac{R}{\mathcal{S}_{0}}10\log_{10}2
\end{eqnarray*}
Examples in \cite{Ver02IT} show that this is a good approximation
for many channels up to fairly high spectral efficiencies, e.g., 1
bit/s/Hz. Further, \cite{Ver02IT} shows that $\ebnomin$ and $\slope_{0}$
can be determined by the first and second order Taylor expansion coefficients
of $R\left(\snr\right)$ at $\snr=0$, namely
\begin{eqnarray}
\ebnomin & = & \frac{\log_{e}2}{\dot{R}\left(0\right)}\label{eq:d1}\\
\slope_{0} & = & -\frac{2\left(\dot{R}\left(0\right)\right)^{2}}{\ddot{R}\left(0\right)}\label{eq:d2}
\end{eqnarray}
where $\dot{R}\left(0\right)=\left.\frac{dR\left(\snr\right)}{d\snr}\right|_{\snr=0}$
and $\ddot{R}\left(0\right)=\left.\frac{d^{2}R\left(\snr\right)}{d\snr^{2}}\right|_{\snr=0}$. 

The reference point for wideband slope is the point-to-point AWGN
(additive white Gaussian noise) channel, which has a wideband slope
of 2. The wideband slope also characterizes the bandwidth required
to transmit at a given rate (in the low-$\mathrm{SNR}$ regime). For
example, if the wideband slope is decreased from 2 to 1, twice the
bandwidth is required for transmitting at a given rate.

The wideband slope for interference channels was considered for the
2-user channel in \cite{CaireVerduAl04} (a generalization to QPSK
can be found in \cite{WieseAl10}). They showed that TDMA (time-division
multiple access) is not efficient in the low-$\mathrm{SNR}$ regime.
In Section \ref{TwoUser.sec} we will extend the results of \cite{CaireVerduAl04}.
However, the main focus of the paper is the $K$-user channel, and
in particular how interference alignment as in \cite{CadambeJafar07}
can be used in the low-$\mathrm{SNR}$ regime.

Traditional interference alignment as in \cite{CadambeJafar07} does
not work in the low-$\mathrm{SNR}$ regime. The results in \cite{CadambeJafar07}
depend on time or frequency selectivity of the channel. However, to
achieve the minimum energy per bit in a non-flat channel, all data
needs to be transmitted on the strongest channel only -- which means
that the wideband slope is poor (e.g., $\frac{2}{K}$ for a $K$-user
interference channel if only the strongest user transmits). On the
other hand, delay differences between different paths can be effectively
used. Delay differences for interference alignment was also considered
in \cite{CadJaf07ACSSC,GrokTseYat08,MaBo09}. However, delay is a
more natural fit for the low-$\mathrm{SNR}$ regime. Namely, as the
bandwidth $B\to\infty$ even the smallest delay will eventually be
magnified to the point of being much larger than the symbol duration.
Therefore, delays can be efficiently manipulated and used for high
bandwidth.

In this paper we will prove that interference alignment using delays
can be used to achieve half the wideband slope of an interference-free
channel, similar to losing half the degrees of freedom in the high-SNR
regime. We will also show that generally it is difficult to obtain
a larger wideband slope. The fact that wideband slope is reduced by
only half means that near single-user performance can be obtained
in the low-power regime. For example, if it is desired to transmit
at $R=0.5$ spectral efficiency, in the interference-free channel
this requires $0.6\mathrm{dB}$ extra energy over the minimum energy
per bit for $R=0$. With interference, $1.2\mathrm{dB}$, e.g., $0.6\mathrm{dB}$
extra energy is needed to overcome interference, \emph{independently
of the number of users}.

\section{System Model\label{systemmodel.sec}}

We consider a scalar complex $K$-user interference channel with Gaussian
noise with line-of-sight (LOS) propagation. There are $K$ transmitters,
numbered 1 to $K$, and $K$ receivers, also numbered 1 to $K$. Transmitter
$j$ needs to transmit a message to receiver $j$, and receiver $j$
has no need for messages from transmitter $i,\, i\neq j$. All transmitters
and receivers have one antenna. As the specifics of the wireless model
affect the results, we will discuss in more details the physical modeling
of the system. The transmitters and receivers are placed in a two
or three dimensional space, where the distance from transmitter $i$
to receiver $j$ is denoted $d_{ji}$. Consistent with the LOS model,
we assume the wireless signal propagates $ $\emph{directly} from
transmitter $j$ to receiver $i$, and the delay in signal arrival
is therefore determined by $d_{ji}$.

While the LOS model is particular, it does apply directly to some
real systems, for example fractionated spacecraft \cite{F6}. An extension
of results to multipath may be possible, but far from straightforward.
Therefore, to obtain a concise mathematical theory we restrict attention
to the LOS model. 

Consider at first a single transmitter-receiver pair, $i$ and $j$.
Let the complex discrete-time transmitted signal of transmitter $i$
be $x_{i}[n]$ and the corresponding baseband (continuous-time) signal
be $x_{i}(t)$ with (two-sided) bandwidth $B$. Specifically, to satisfy
a strict band limit we must have
\begin{eqnarray*}
x_{i}\left(t\right) & = & \sum_{n}x_{i}\left[n\right]\sinc\left(Bt-n\right).
\end{eqnarray*}
This is modulated with the carrier signal $c(t)=\exp\iota(\omega_{0}(t-\varsigma_{i}))$,
where $\omega_{0}$ is the carrier frequency and $\varsigma_{i}$
is the delay (phase offset) in the oscillator at transmitter $i$
(and $\iota=\sqrt{-1}$). The real part is transmitted,
\begin{eqnarray*}
s_{i}(t) & = & \Re\left\{ \exp\iota(\omega_{0}(t-\varsigma_{i}))x_{i}(t)\right\} \\
 & = & \cos(\omega_{0}(t-\varsigma_{i}))\Re\{x_{i}(t)\}-\sin(\omega_{0}(t-\varsigma_{i}))\Im\{x_{i}(t)\}.
\end{eqnarray*}
The received signal at receiver $j$ is
\begin{eqnarray*}
r_{j}(t) & = & A_{ji}\Re\{\exp\iota(\omega_{0}(t-\varsigma_{i}-\tau_{ji}))x_{i}(t-\tau_{ji})\}+\tilde{z}_{j}(t)\\
\tau_{ji} & = & \frac{d_{ji}}{c}
\end{eqnarray*}
where $A_{ji}$ is an attenuation factor, $c$ is the speed of light,
and $\tilde{z}_{j}(t)$ is white Gaussian noise with power spectral
density $N_{0}$. This is modulated to baseband by multiplying with
$\exp(-\iota\omega_{0}(t-\upsilon_{j}))$, where $\upsilon_{j}$ is
the delay in the oscillator at receiver $j$, and using a lowpass
filter, resulting in the baseband signal
\begin{eqnarray*}
y_{j}(t) & = & A_{ji}\exp(\iota\omega_{0}(\varsigma_{i}+\tau_{ji}-\upsilon_{j}))x_{i}(t-\tau_{ji})+z_{j}(t).
\end{eqnarray*}
This expression is valid on the assumption that $\omega_{0}>B$. Here
$z_{j}(t)$ is white Gaussian noise filtered to a bandwidth $B$. 

Return now to the interference channel. When all users transmit, the
received signal at receiver $j$ is
\begin{eqnarray}
y_{j}(t) & = & A_{jj}\exp(\iota\omega_{0}(\varsigma_{j}+\tau_{jj}-\upsilon_{j}))x_{j}(t-\tau_{jj})+\sum_{i\neq j}A_{ji}\exp(\iota\omega_{0}(\varsigma_{i}+\tau_{ji}-\upsilon_{j}))x_{i}(t-\tau_{ji})+z_{j}(t).\label{eq:continuous channel}
\end{eqnarray}
This is sampled at the Nyquist frequency $f_{s}=B$ (as $B$ is the
two-sided bandwidth). Let 
\begin{eqnarray}
n_{ji} & = & \left\lfloor \tau_{ji}B+{\textstyle \frac{1}{2}}\right\rfloor \\
\delta_{ji} & = & \tau_{ji}B-\left\lfloor \tau_{ji}B+{\textstyle \frac{1}{2}}\right\rfloor \label{fracdelay.eq}
\end{eqnarray}
where $\left\lfloor x\right\rfloor $ is the largest integer smaller
than or equal to $x$. Without loss of generality we can assume that
the received signal at receiver $j$ is sampled symbol-synchronous
with the desired signal. Then the discrete-time model is
\begin{eqnarray}
y_{j}[n] & = & A_{jj}\exp(\iota\omega_{0}(\varsigma_{j}+\tau_{jj}-\upsilon_{j}))x_{j}[n-n_{jj}]+\sum_{i\neq j}A_{ji}\exp(\iota\omega_{0}(\varsigma_{i}+\tau_{ji}-\upsilon_{j}))\tilde{x}_{i}[n-n_{ji}]+z_{j}[n]\label{eq:discrete 1}
\end{eqnarray}
where $z_{j}[n]$ is as sequence of i.i.d circularly symmetric random
variables, $z_{j}[n]\sim\mathcal{N}\left(0,BN_{0}\right)$, and
\begin{eqnarray}
\tilde{x}_{i}[n] & = & \sum_{m=-\infty}^{\infty}x_{i}[m]\sinc(n-m+\delta_{ji}).\label{tildex.eq}
\end{eqnarray}
We will also occasionally make the dependency on the fractional delay
explicit as follows
\begin{eqnarray}
\tilde{x}_{i}[n,\delta_{ji}] & = & \sum_{m=-\infty}^{\infty}x_{i}[m]\sinc(n-m+\delta_{ji}).\label{eq:delayed}
\end{eqnarray}
By the Shannon sampling theorem, this discrete-time model is equivalent
with the original continuous-time model. Results do not change if
we normalize the time at each receiver so that $n_{jj}=0$. And as
the carrier frequency is large, the phases $\exp(\iota\omega_{0}(\varsigma_{i}+\tau_{ji}-\upsilon_{j}))$
can be reasonably modeled as independent uniform random variables
$\theta_{ji}$ over the unit circle. We therefore arrive at the following
expression for the received signal
\begin{eqnarray}
y_{j}[n] & = & A_{jj}\exp\iota\theta_{jj}x_{i}[n]+\sum_{i\neq j}A_{ji}\exp\iota\theta_{ji}\tilde{x}_{i}[n-n_{ji}]+z_{j}[n]\nonumber \\
 & = & C_{jj}x_{j}[n]+\sum_{i\neq j}C_{ji}\tilde{x}_{i}[n-n_{ji}]+z_{j}[n]\label{eq:discrete channel}
\end{eqnarray}
where $C_{ji}=A_{ji}\exp\iota\theta_{ji}$.

Notice that this model makes no assumptions on or approximations of
modulation, e.g., it does not assume rectangular waveforms. Transmission
in our model is \emph{strictly} bandlimited to a bandwidth $B$, as
opposed to \cite{CadJaf07ACSSC}.

\subsection{\label{sub:Approaching-the-Low-}Approaching the Low-$\snr$ Regime:
Large $B$ Case and Small $B$ Case }

What is interesting is that there are two distinct ways to approach
the low-$\snr$ regime, which have very different impacts on the performance
of the interference channel defined by (\ref{eq:discrete channel}).
Although approaching the low-$\snr$ regime by letting $B\rightarrow\infty$
is emphasized in previous papers, it is not the only way. As can be
noted from the definition of SNR (\ref{eq:def snr}), SNR approaches
zero if either $B\to\infty$ or $P\to0$. Consider a point-to-point
AWGN channel with spectral efficiency
\begin{eqnarray*}
R & = & \log\left(1+\frac{P}{BN_{0}}\right).
\end{eqnarray*}
The low-$\snr$ results are based on a Taylor series of $\log(1+x)$
, as also seen by (\ref{eq:d1}-\ref{eq:d2}); therefore as long as
$\snr=\frac{P}{BN_{0}}\to0$ low-power results such as minimum energy
per bit and wideband slope are valid. The key is that the spectral
efficiency $R\to0$, not that $B\to\infty$. For the interference
channel (\ref{eq:discrete channel}), on the other hand, different
results are obtained depending on how the low-$\snr$ regime is approached.

In the first approach, let $B\to\infty$ while $P$ is fixed and finite.
We call this approach the \emph{large bandwidth case}. In this case,
the propagation delay is large compared with the symbol duration,
i.e., as $B\to\infty$, $n_{ji}$ can become arbitrarily large even
for very small $\tau_{ji}$. 

In the second approach, let $P\rightarrow0$ while $B$ is fixed and
finite. We further assume that the propagation delay is much smaller
than the symbol duration, i.e., $\tau_{ij}B\ll1$. Under this assumption,
$n_{ji}=0\;\mathrm{and}\;\delta_{ji}\approx0$. This approach is called
the \emph{small bandwidth case}. 

The large bandwidth case is the topic of this paper; the small bandwidth
case will be considered in a later paper.

\subsection{Performance criteria}

In \cite{CaireVerduAl04} the whole slope region of the interference
region in the 2-user case was analyzed. However, for more than two
users it is complicated to compare complete slope regions, and we
are therefore looking at a single quantity to characterize performance.
We denote the power constraint for each user $P_{i}$ and the spectral
efficiency $R_{i}$; we further set $\snr_{i}=\frac{P_{i}}{N_{0}B}$.
We consider two different constraints
\begin{itemize}
\item \emph{The equal power} \emph{constraint}. In this case we maximize
the sum spectral efficiency $R_{s}=R_{1}+R_{2}+\cdots R_{K}$ under
the constraint $P_{1}=P_{2}=\cdots=P_{K}$, i.e., $\snr_{1}=\snr_{2}=\cdots=\snr_{K}$.
We want to characterize the wideband slope of the sum spectral efficiency
$R_{s}$.
\item \emph{The equal rate constraint}. In this case we minimize the total
power per Hz $\snr=\snr_{1}+\snr_{2}+\cdots+\snr_{K}$ under the constraint
$R=R_{1}=R_{2}=\cdots=R_{K}$. We want to characterize the wideband
slope of the sum spectral efficiency $R_{s}=K\cdot R$.
\end{itemize}

The equal power constraint could correspond to a scenario where each
user needs to consume energy at the same rate, e.g., so that batteries
last the same for all users. The equal rate constraint could correspond
to a scenario where we want to minimize total system energy consumption.
Each constraint can be easily generalized to unbalanced cases, e.g.
$\mu_{1}\snr_{1}=\mu_{2}\snr_{2}=\cdots=\mu_{K}\snr_{K}$, but we
only consider the balanced case here to keep results concise.

As performance measure we use
\begin{eqnarray*}
\Delta\mathcal{S}_{0} & = & \frac{\mathcal{S}_{0}}{\mathcal{S}_{0,\mbox{no interference}}}.
\end{eqnarray*}
The quantity $\mathcal{S}_{0,\mbox{no inteference}}$ is the wideband
slope of a $K$-user interference channel where all interference links
are nulled, $|C_{ij}|=0$, $i\neq j$, but the direct links $C_{ii}$
are unchanged. We can interpret $\Delta\mathcal{S}_{0}$ as the loss
in wideband slope due to interference, or equivalently $\left(\Delta\mathcal{S}_{0}\right)^{-1}$
as (approximately) the additional bandwidth required to overcome interference.
Alternatively, if we define
\begin{eqnarray*}
\Delta E_{b} & = & 10\log_{10}\ebno-10\log_{10}\ebno_{\min}
\end{eqnarray*}
as the extra energy required to operate at a spectral efficiency $R>0$,
we have
\begin{eqnarray*}
\Delta E_{b} & \approx & \left(\Delta\mathcal{S}_{0}\right)^{-1}\Delta E_{b,\mbox{no interference}}
\end{eqnarray*}
for small increases in spectral efficiency. Thus, $\left(\Delta\mathcal{S}_{0}\right)^{-1}$
also measures the amount of energy needed to overcome interference.

\subsection{$\ebnomin$ of the Interference Channel}

The papers \cite{Ver02IT} and \cite{CaiTunVer04IT} show that the
minimum energy per bit of the interference channel is equal to that
of an interference-free channel, achievable by Treating Interference
as Noise (TIN) and TDMA. The following theorem gives the transmitted
$\ebnomin$ under the two different constraints.
\begin{thm}
\label{thm: correct ebnomin}The minimum energy per bit of the interference
channel defined by (\ref{eq:continuous channel}) is 
\begin{eqnarray}
\frac{E_{b}}{N_{0}}_{\mathrm{min}} & = & \frac{\sum\left(\left|C_{jj}\right|^{-2}\right)}{K}\log_{e}2\label{eq:ebnomin equal rate}
\end{eqnarray}
under the equal rate constraint; and 
\begin{eqnarray}
\frac{E_{b}}{N_{0}}_{\mathrm{min}} & = & \frac{K\log_{e}2}{\sum_{j=1}^{K}\left|C_{jj}\right|^{2}}\label{eq:ebnomin equal power}
\end{eqnarray}
under the equal power constraint.
\end{thm}
The best known achievable rate for the interference channel is the
Han-Kobayashi region \cite{HanKobayashi81}. For the Gaussian interference
channel, in particular the idea of transmitting common messages has
been shown to be powerful \cite{EtkinTseWang07}. However, the common
message does have a higher $\ebno$ in the low power limit than the
minimum, and therefore does not improve the wideband slope. To make
fair comparison any bound imposed on the wideband slope must have
the correct $\ebnomin$. We emphasize this requirement in the following
remark.
\begin{rem}
\emph{\label{rem: correct ebnomin}For a bound on rate to be useful
as a bound on wideband slope, it needs to have the correct $ $$\ebnomin$
given in Theorem \ref{thm: correct ebnomin}. }
\end{rem}

\section{\label{TwoUser.sec}The 2-User Case}

We will start by analyzing the 2-user case as this is instrumental
for the $K$-user case. As we have discussed in Section \ref{sub:Approaching-the-Low-},
the essential difference between large bandwidth case and small bandwidth
case is that they impact the behavior of propagation delays differently.
It turns out that all results in the 2-user case are independent of
delay, thus independent of how the low-$\snr$ regime is approached.
This indicates that the capacity region in the 2-user case could be
independent of delay in general, but we have not been able to prove
so.

\subsection{Achievable Schemes\label{sub:2-user Achievable-Scheme}}

First we will outline the strategies that can be used for the achievable
rate. In order to use these to inner bound the the sum slope, as mentioned
in Remark \ref{rem: correct ebnomin}, they must have the correct
$\ebnomin$, and that only leaves three strategies
\begin{enumerate}
\item \emph{Interference decoding}.

If $|C_{ji}|>|C_{ii}|$, user $j$ can decode the message from user
$i$, and the capacity region of the interference channel is equivalent
to the capacity region of the multi-access channel formed by transmitter
$i$, transmitter $j$, and receiver $j$, which is 
\begin{eqnarray}
R_{j} & \leq & \log\left(1+\left|C_{jj}\right|^{2}\snr_{j}\right)\label{eq:macj}\\
R_{i} & \leq & \log\left(1+\left|C_{ii}\right|^{2}\snr_{i}\right)\label{eq:maci}\\
R_{i}+R_{j} & \leq & \log\left(1+\left|C_{ji}\right|^{2}\snr_{i}+\left|C_{jj}\right|^{2}\snr_{j}\right).\label{eq:macsum}
\end{eqnarray}
In the low-$\snr$ regime, as $\snr\rightarrow0$, there always exists
some real number $\epsilon>0$ such that if $\snr_{j},\,\snr_{i}<\epsilon$
the sum slope outer bound given by the summation of (\ref{eq:macj})
and (\ref{eq:maci}) is less than the sum slope outer bound given
by (\ref{eq:macsum}) because $|C_{ji}|>|C_{ii}|$. Therefore, (\ref{eq:macsum})
can be discarded and the multi-access bound is equivalent to the rectangular
capacity region of a channel with no interference. Thus, interference
does not affect wideband slope in this case.

\item \emph{Treating interference as noise} (TIN).

The transmitters use i.i.d Gaussian code books, and the receivers
treat the interference as part of the background noise. Notice that
delay does not affect the distribution of interference as $\tilde{x}_{i}[n-n_{ji}]$
has same distribution as $x_{i}[n]$. The achievable $\left(R_{1},R_{2}\right)$
is
\begin{eqnarray}
R_{1} & \leq & \log\left(1+\frac{\left|C_{11}\right|^{2}\snr_{1}}{1+\left|C_{12}\right|^{2}\snr_{2}}\right)\\
R_{2} & \leq & \log\left(1+\frac{\left|C_{22}\right|^{2}\snr_{2}}{1+\left|C_{21}\right|^{2}\snr_{1}}\right).
\end{eqnarray}

\item \emph{TDMA}.

In time-division multiple access the transmitters use orthogonal time
slots. Because of the delay differences, users have to insert buffers
with no transmission around each TDMA frame, so that they are orthogonal
at both users. However, the length of these buffers is finite, so
as the code length converges towards infinity (as required by capacity
analysis), the effect of these buffers on spectral efficiency will
converge towards zero. TDMA therefore achieves the following spectral
efficiency also in the case of delays, 
\begin{eqnarray}
R_{1} & \leq & \frac{1}{2}\log\left(1+2\left|C_{11}\right|^{2}\snr_{1}\right)\\
R_{2} & \leq & \frac{1}{2}\log\left(1+2\left|C_{22}\right|^{2}\snr_{2}\right).\label{eq:tdma2}
\end{eqnarray}

\end{enumerate}
The achivable sum slope can easily be straightforwardly calculated
from these equations using (\ref{eq:d2}). The expressions are too
complex to give much insight, so we will only state them for later
reference for a canonical 2-user channel with symmetric interference
link gains.
\begin{thm}
\label{thm:2-user lower bound}Consider a 2-user interference channel
where $\left|C_{jj}\right|^{2}=1$ and $\left|C_{ji}\right|^{2}=a,\; i\neq j$.
The sum slope is inner bounded by 

\begin{eqnarray}
\slope_{0} & \geq & \left\{ \begin{array}{ll}
4 & a>1\\
2 & \frac{1}{2}<a<1\\
\frac{4}{1+2a} & a\leq\frac{1}{2}
\end{array}\right.\label{eq:sum slope 2-user equal power}
\end{eqnarray}
under both the equal power constraint and the equal rate constraint.
\end{thm}

\subsection{\label{sub:2-user Outer-Bounds}Outer Bounds}

In this section, we will state some sum slope outer bounds, and discuss
the so-called noisy interference channel, where the exact sum slope
is known.

The following theorem generalizes Theorem 2 from \cite{Kra04IT} to
channels with delay.
\begin{thm}[Kramer's bound]
\label{thm:Kramer}Suppose that $|C_{21}|<|C_{11}|$. Then
\begin{eqnarray}
R_{1} & \leq & \log\left(1+\left|C_{11}\right|^{2}\snr_{1}\right)\label{kramerbound.eq}\\
R_{2} & \leq & \log\left(\frac{1+\left|C_{22}\right|^{2}\snr_{2}+\left|C_{21}\right|^{2}\snr_{1}}{\frac{\left|C_{21}\right|^{2}}{\left|C_{11}\right|^{2}}2^{R_{1}}+1-\frac{\left|C_{21}\right|^{2}}{\left|C_{11}\right|^{2}}}\right)\label{eq:kramerbound2}\\
R_{1}+R_{2} & \leq & \log\left(1+\left|C_{11}\right|^{2}\snr_{1}\right)\\
 &  & +\log\left(1+\frac{\left|C_{22}\right|^{2}\snr_{2}}{1+\left|C_{21}\right|^{2}\snr_{1}}\right)\label{eq:kramer sum alt}
\end{eqnarray}
independent of delay. \end{thm}
\begin{IEEEproof}
Put $C_{12}=0$ to enlarge the capacity region. Now assume that, different
from the system model (\ref{eq:discrete channel}), receiver 2 also
samples the received signal synchronously with the transmitted signal
of user 1%
\footnote{According to the sampling theorem, this does not change the capacity
region.%
}. A Z-channel with delay is formed: 
\begin{eqnarray}
y_{1}^{\prime}[n] & = & C_{11}x_{1}[n]+z_{1}[n]\nonumber \\
y_{2}[n] & = & C_{22}\tilde{x}_{2}[n-n_{22}]+C_{21}x_{1}[n]+z_{2}[n]\label{eq:z y2}
\end{eqnarray}
where $\tilde{x}_{2}[n]$ is defined by (\ref{tildex.eq}). 

Next, we show that the capacity region of (\ref{eq:z y2}) is independent
of delay. The channel (a) and (b) illustrated in Figure \ref{fig:degrade channel}
have identical capacity regions because $p\left(\left.\check{y}_{2}\right|x_{1},\tilde{x}_{2}\right)$
and $p\left(\left.y_{2}\right|x_{1},\tilde{x}_{2}\right)$ have the
same distribution. $\check{z}_{2}[n]$ is i.i.d Gaussian noise independent
of $z_{1}[n]$ and the input signals, with power $\left(1-\frac{|C_{21}|^{2}}{|C_{11}|^{2}}\right)N_{0}B$.
Because $\frac{|C_{21}|^{2}}{|C_{11}|^{2}}<1$, such $\check{z}_{2}[n]$
is guaranteed to exist. The argument is identical to (a)\textasciitilde{}(c)
of Figure 6 in \cite{Cos85IT}. Details are skipped here.

\begin{figure}[tbh]
\centering{}\includegraphics{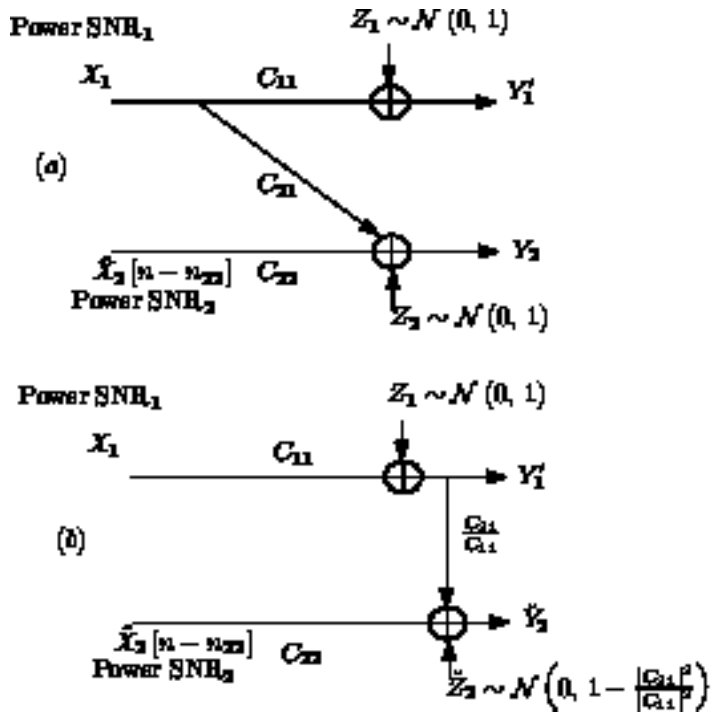}\caption{\label{fig:degrade channel}Channels with Equivalent Capacity Region}
\end{figure}
The channel $\left(b\right)$ has the form 
\begin{eqnarray}
y_{1}^{\prime}[n] & = & C_{11}x_{1}[n]+z_{1}[n]\label{eq:degrade 1}\\
\check{y}_{2}[n] & = & C_{22}\tilde{x}_{2}[n-n_{22}]+\frac{C_{21}}{C_{11}}y_{1}^{\prime}[n]+\check{z}_{2}[n].\label{eq:degrade 2}
\end{eqnarray}
Using Fano's inequality as usual, we can now bound the capacity of
this channel by 
\begin{eqnarray}
nR_{2}-n\epsilon_{n} & \leq & h(\check{y}_{2}^{n})-h(\check{y}_{2}^{n}|\tilde{x}_{2})\label{eq:data pr1}\\
 & \leq & h(\check{y}_{2}^{n})-h(\check{y}_{2}^{n}|w_{2})\label{eq:data pr2}\\
 & = & h(\check{y}_{2}^{n})-h\left(\frac{C_{21}}{C_{11}}y_{1}^{\prime n}+\check{z}_{2}^{n}|w_{2}\right),\label{kramerproof.eq}
\end{eqnarray}
where $w_{2}$ is the message sent by transmitter 2. The step (\ref{eq:data pr1})
to (\ref{eq:degrade 2}) is from the data processing inequality, as
$\tilde{x}_{2}$ is a function of the transmitted codeword $x_{2}$,
which is a function of $w_{2}$. The second term in (\ref{kramerproof.eq})
is independent of delay, and can be lower bounded by the entropy power
inequality \cite{Bergmans74}. The first term can be upper bounded
by the delay-free case. Therefore, the capacity region of (\ref{eq:z y2})
is identical to that of the channel without delay. The papers \cite{Cos85IT}
and \cite{Kra04IT} show that the capacity region of delay-free channel
can be derived from an equivalent degraded broadcast channel. Given
Theorem 1 in \cite{ahm05ISIT} its rate region has upper bound (\ref{kramerbound.eq}).
Finally, it is easy to see that the capacity region of (\ref{eq:discrete channel})
is contained within that of the Z-channel. The equation (\ref{eq:kramer sum alt})
is a restatement of (47) in \cite{Kra04IT}.
\end{IEEEproof}
We use Theorem \ref{thm:Kramer} to obtain a sum slope outer bound
under the equal power constraint as follows,
\begin{cor}
\label{cor:2-user outer bound}Suppose that $|C_{21}|<|C_{11}|$.
Under the equal power constraint, the wideband slope for the sum rate
has outer bound
\begin{eqnarray}
\mathcal{S}_{0} & \leq & 2\frac{(|C_{11}|^{2}+|C_{22}|^{2})^{2}}{2|C_{21}|^{2}|C_{22}|^{2}+|C_{11}|^{4}+|C_{22}|^{4}}\label{eq:coro2 1}\\
\Delta\mathcal{S}_{0} & \leq & \frac{1}{2\frac{|C_{21}|^{2}|C_{22}|^{2}}{|C_{11}|^{4}+|C_{22}|^{4}}+1},\label{eq:coro2}
\end{eqnarray}
independent of delay. \end{cor}
\begin{IEEEproof}
This result can be easily shown combining (\ref{eq:kramer sum alt})
and the formulas (\ref{eq:d1}) and (\ref{eq:d2}).
\end{IEEEproof}
Results similar to Theorem \ref{thm:Kramer} and Corollary \ref{cor:2-user outer bound}
can be obtained for $\left|C_{12}\right|<\left|C_{22}\right|$ case
by interchanging the indices $'1'$ and $'2'$.

\begin{figure}[tbh]
\begin{centering}
\includegraphics[width=3.5in]{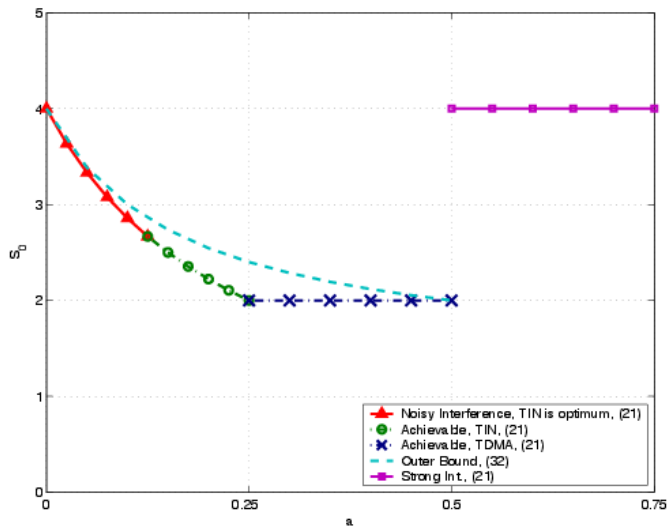}
\par\end{centering}

\caption{Sum slope versus $\frac{|C_{21}|}{|C_{11}|}$. In the legend, TIN
stands for treating interference as noise. \label{twouser.fig}}
\end{figure}

For the equal-rate constraint if only one interference link is weak,
bound (\ref{kramerbound.eq}) does not have the correct \emph{$\ebnomin$
and therefore cannot be used for bounding the wideband slope }by Remark
\ref{rem: correct ebnomin}\emph{.} If both interference links are
weak, we have following corollary.
\begin{cor}
\label{cor:kramer eqrate}Suppose that $|C_{21}|<|C_{11}|$ and $|C_{12}|<|C_{22}|$.
Under the equal rate constraint, the wideband slope for the sum rate
is upper bounded by
\begin{eqnarray*}
\mathcal{S}_{0} & \leq & 4\cdot\left(|C_{11}|^{2}+|C_{22}|^{2}\right)\left(1-\frac{\left|C_{12}\right|^{2}}{\left|C_{22}\right|^{2}}\frac{\left|C_{21}\right|^{2}}{\left|C_{11}\right|^{2}}\right)\cdot\left(|C_{11}|^{2}+|C_{22}|^{2}+\right.\\
 &  & \left.|C_{21}|^{2}\left(2-3\frac{\left|C_{12}\right|^{2}}{\left|C_{22}\right|^{2}}\right)+|C_{12}|^{2}\left(2-3\frac{\left|C_{21}\right|^{2}}{\left|C_{11}\right|^{2}}\right)\right)^{-1}\\
\Delta\mathcal{S}_{0} & \leq & \left(|C_{11}|^{2}+|C_{22}|^{2}\right)\left(1-\frac{\left|C_{12}\right|^{2}}{\left|C_{22}\right|^{2}}\frac{\left|C_{21}\right|^{2}}{\left|C_{11}\right|^{2}}\right)\cdot\left(|C_{11}|^{2}+|C_{22}|^{2}+\right.\\
 &  & \left.|C_{21}|^{2}\left(2-3\frac{\left|C_{12}\right|^{2}}{\left|C_{22}\right|^{2}}\right)+|C_{12}|^{2}\left(2-3\frac{\left|C_{21}\right|^{2}}{\left|C_{11}\right|^{2}}\right)\right)^{-1}
\end{eqnarray*}
independent of delay. \end{cor}
\begin{IEEEproof}
(\ref{kramerbound.eq}) gives 
\begin{eqnarray}
\left|C_{21}\right|^{2}\snr_{1}+\left|C_{22}\right|^{2}\snr_{2} & \geq & 2^{R_{2}}\left(\frac{\left|C_{21}\right|^{2}}{\left|C_{11}\right|^{2}}2^{R_{1}}-\frac{\left|C_{21}\right|^{2}}{\left|C_{11}\right|^{2}}+1\right)-1\label{2user eqrate 1.eq}\\
\left|C_{11}\right|^{2}\snr_{1}+\left|C_{12}\right|^{2}\snr_{2} & \geq & 2^{R_{1}}\left(\frac{\left|C_{12}\right|^{2}}{\left|C_{22}\right|^{2}}2^{R_{2}}-\frac{\left|C_{12}\right|^{2}}{\left|C_{22}\right|^{2}}+1\right)-1.\label{2user eqrate 2.eq}
\end{eqnarray}
Under the equal rate constraint, $R_{1}=R_{2}=\frac{R_{s}}{2}$, and
our objective is to minimize $\snr_{1}+\snr_{2}$. We construct the
following optimization problem
\begin{eqnarray*}
 & \min & \snr_{1}+\snr_{2}\\
 & \mathrm{s.t.} & \mathbf{A}\left(\begin{array}{c}
\snr_{1}\\
\snr_{2}
\end{array}\right)\geq\mathrm{b}\\
 &  & P_{j}\geq0
\end{eqnarray*}
where $\mathbf{A}=\left(\begin{array}{cc}
\left|C_{21}\right|^{2} & \left|C_{22}\right|^{2}\\
\left|C_{11}\right|^{2} & \left|C_{12}\right|^{2}
\end{array}\right)$ and $\mathrm{b}=\left(\begin{array}{c}
2^{\nicefrac{R_{s}}{2}}\left(\frac{\left|C_{21}\right|^{2}}{\left|C_{11}\right|^{2}}2^{\nicefrac{R_{s}}{2}}-\frac{\left|C_{21}\right|^{2}}{\left|C_{11}\right|^{2}}+1\right)-1\\
2^{\nicefrac{R_{s}}{2}}\left(\frac{\left|C_{12}\right|^{2}}{\left|C_{22}\right|^{2}}2^{\nicefrac{R_{s}}{2}}-\frac{\left|C_{12}\right|^{2}}{\left|C_{22}\right|^{2}}+1\right)-1
\end{array}\right)$. Using simple linear programming principles, one optimal solution
can be found at the vertex of the feasible region. That is, $\left.\snr_{1}+\snr_{2}\right|_{\min}=\snr_{1o}+\snr_{2o}$
where$\left(\begin{array}{c}
\snr_{1o}\\
\snr_{2o}
\end{array}\right)=\mathbf{A}^{-1}\mathrm{b}>0$. We solve this simple linear system and get 
\begin{eqnarray}
\snr_{1o} & = & \left|C_{11}\right|^{-2}\cdot\frac{2^{\frac{R_{s}}{K}}\left(2^{\frac{R_{s}}{K}}-1\right)\frac{\left|C_{12}\right|^{2}}{\left|C_{22}\right|^{2}}\left(1-\frac{\left|C_{21}\right|^{2}}{\left|C_{11}\right|^{2}}\right)+\left(1-\frac{\left|C_{21}\right|^{2}}{\left|C_{11}\right|^{2}}\right)\left(2^{\frac{R_{s}}{K}}-1\right)}{1-\frac{\left|C_{12}\right|^{2}}{\left|C_{22}\right|^{2}}\frac{\left|C_{21}\right|^{2}}{\left|C_{11}\right|^{2}}}\label{2user eqrate 3.eq}\\
\snr_{2o} & = & \left|C_{22}\right|^{-2}\cdot\frac{2^{\frac{R_{s}}{K}}\left(2^{\frac{R_{s}}{K}}-1\right)\frac{\left|C_{21}\right|^{2}}{\left|C_{11}\right|^{2}}\left(1-\frac{\left|C_{12}\right|^{2}}{\left|C_{22}\right|^{2}}\right)+\left(1-\frac{\left|C_{12}\right|^{2}}{\left|C_{22}\right|^{2}}\right)\left(2^{\frac{R_{s}}{K}}-1\right)}{1-\frac{\left|C_{12}\right|^{2}}{\left|C_{22}\right|^{2}}\frac{\left|C_{21}\right|^{2}}{\left|C_{11}\right|^{2}}}.\label{2user eqrate 4.eq}
\end{eqnarray}
Now we have the expression of sum power as a function of sum rate.
The following formulas are equivalent to (\ref{eq:d1}) and (\ref{eq:d2}).
\begin{eqnarray}
\frac{E_{b}}{N_{0}}_{\mathrm{min}} & = & \left.\frac{d\snr\left(R\right)}{dR}\right|_{R=0}\label{eq:alt1}\\
\mathcal{S} & = & \frac{2\left.\frac{d\snr\left(R\right)}{dR}\right|_{R=0}}{\left.\frac{d^{2}\snr\left(R\right)}{dR^{2}}\right|_{R=0}}\log2.\label{eq:alt2}
\end{eqnarray}
They can be proved using a technique similar to (140)\textasciitilde{}(144)
in \cite{Ver02IT}. Details are skipped here. Combining (\ref{2user eqrate 3.eq}),
(\ref{2user eqrate 4.eq}) and (\ref{eq:alt2}), we have
\begin{eqnarray*}
\frac{E_{b}}{N_{0}}_{\mathrm{min}} & = & \frac{\left(\left|C_{11}\right|^{-2}+\left|C_{22}\right|^{-2}\right)}{2}\log_{e}2\\
\mathcal{S}_{0} & = & 4\cdot\left(|C_{11}|^{2}+|C_{22}|^{2}\right)\left(1-\frac{\left|C_{12}\right|^{2}}{\left|C_{22}\right|^{2}}\frac{\left|C_{21}\right|^{2}}{\left|C_{11}\right|^{2}}\right)\cdot\left(|C_{11}|^{2}+|C_{22}|^{2}\right.\\
 &  & \left.+|C_{21}|^{2}\left(2-3\frac{\left|C_{12}\right|^{2}}{\left|C_{22}\right|^{2}}\right)+|C_{12}|^{2}\left(2-3\frac{\left|C_{21}\right|^{2}}{\left|C_{11}\right|^{2}}\right)\right)^{-1}
\end{eqnarray*}

\end{IEEEproof}
For the 2-user interference channel without delay, \cite{AnnVee08},
\cite{ShaKraChe07IT} and \cite{MotKha08IT} show that there exists
a class of channels whose optimal sum spectral efficiency can be achieved
by i.i.d. Gaussian inputs and treating interference as noise. This
class of channel is one of the few where the exact capacity is known,
and consequently also the exact sum slope.We here extend these results
to channels with delay.
\begin{thm}
\label{thm:noisy with delay}For a 2-user interference channel defined
by (\ref{eq:discrete channel}), if there exist complex numbers $\rho_{1}$,$\rho_{2}$
and positive real numbers $\sigma_{1}^{2}$, $\sigma_{2}^{2}$ such
that, 
\begin{align}
\left|\rho_{1}\right|^{2} & \leq\sigma_{1}^{2}\leq1-\frac{\left|\rho_{2}\right|^{2}}{\sigma_{2}^{2}}\label{eq:noisy cond 1}\\
\left|\rho_{2}\right|^{2} & \leq\sigma_{2}^{2}\leq1-\frac{\left|\rho_{1}\right|^{2}}{\sigma_{1}^{2}}\label{eq:noisy cond 2}\\
C_{21} & =\frac{\rho_{1}C_{11}}{\left|C_{12}\right|^{2}\snr_{2}+1}\label{eq:noisy cond 3}\\
C_{12} & =\frac{\rho_{1}C_{11}}{\left|C_{12}\right|^{2}\snr_{2}+1},\label{eq:noisy cond 4}
\end{align}
then the optimal sum capacity 
\begin{eqnarray}
R_{1}+R_{2} & \leq & \log\left(1+\frac{|C_{11}|^{2}\snr_{1}^{2}}{1+|C_{12}|^{2}\snr_{2}}\right)+\log\left(1+\frac{|C_{22}|^{2}\snr_{2}^{2}}{1+|C_{21}|^{2}\snr_{1}}\right)\label{eq:noisy sum r}
\end{eqnarray}
is achievable by i.i.d. Gaussian input and treating interference as
noise at the receivers. Further, (\ref{eq:noisy cond 1}) \textasciitilde{}
(\ref{eq:noisy cond 4}) are satisfied as long as 
\begin{eqnarray}
\sqrt{\frac{|C_{12}|^{2}}{|C_{22}|^{2}}}\left(1+|C_{21}|^{2}\snr_{1}\right)+\sqrt{\frac{|C_{21}|^{2}}{|C_{11}|^{2}}} & \left(1+|C_{12}|^{2}\snr_{2}\right)\leq & 1.\label{eq:noisy condition}
\end{eqnarray}
. \end{thm}
\begin{IEEEproof}
Please see Appendix \ref{sec:Proof-of-Theorem noisy with delay}. 
\end{IEEEproof}
Theorem \ref{thm:noisy with delay} is identical to the case where
there is no delay, which is discussed in \cite[Theorem 6]{ShaCheKra09}.
We now use Theorem \ref{thm:noisy with delay} to derive the exact
sum slope,
\begin{cor}
\label{cor:low-snr noisy}Consider the 2-user interference channel
defined by (\ref{eq:discrete channel}). Under the equal power constraint,
if the channel coefficients satisfy
\begin{eqnarray}
\sqrt{\frac{|C_{12}|^{2}}{|C_{22}|^{2}}}+\sqrt{\frac{|C_{21}|^{2}}{|C_{11}|^{2}}} & < & 1,\label{TINcond.eq}
\end{eqnarray}
then i.i.d. Gaussian inputs and treating interference as noise achieve
the optimal sum slope $\mathcal{S}_{0}$, which is 
\begin{eqnarray}
\mathcal{S}_{0} & = & \frac{2\left(|C_{11}|^{2}+|C_{22}|^{2}\right)^{2}}{|C_{11}|^{4}+|C_{22}|^{4}+2\left(|C_{11}|^{2}|C_{12}|^{2}+|C_{21}|^{2}|C_{22}|^{2}\right)}\label{TIN.eq}\\
\Delta\mathcal{S}_{0} & = & 1+\frac{2\left(|C_{11}|^{2}|C_{12}|^{2}+|C_{21}|^{2}|C_{22}|^{2}\right)}{|C_{11}|^{4}+|C_{22}|^{4}}.
\end{eqnarray}
\end{cor}
\begin{IEEEproof}
Under the equal power constraint where $\snr_{i}=\frac{\snr_{s}}{2}$
there must exist some $\epsilon>0$, such that if $\snr<\epsilon$
then (\ref{eq:noisy condition}) can be satisfied. Because the low-$\mathrm{SNR}$
regime is approached as $\snr\rightarrow0$, this gives (\ref{TINcond.eq}).
Given (\ref{eq:noisy sum r}), under the equal power constraint the
sum rate achieved by treating interference as noise is

\begin{eqnarray}
R_{s} & \leq & \log\left(1+\frac{|C_{11}|^{2}\frac{\snr}{2}}{1+|C_{12}|^{2}\frac{\snr}{2}}\right)\nonumber \\
 &  & +\log\left(1+\frac{|C_{22}|^{2}\frac{\snr}{2}}{1+|C_{21}|^{2}\frac{\snr}{2}}\right).\label{eq:equal power noisy som r}
\end{eqnarray}
Combining (\ref{eq:equal power noisy som r}) with (\ref{eq:d1})
and (\ref{eq:d2}) we have (\ref{TIN.eq}).
\end{IEEEproof}
Figure \ref{twouser.fig} illustrates the sum slope region of a 2-user
interference channel with unit direct link gain, and symmetric cross
link gain, that is, $\left|C_{11}\right|^{2}=\left|C_{22}\right|^{2}=1$,
and $\left|C_{12}\right|^{2}=\left|C_{21}\right|^{2}=a$. In this
figure, the inner bound is given by Theorem \ref{thm:2-user lower bound}:
the inner bounds labeled {}``Strong Int.'', {}``Achievable, TDMA'',
and {}``Achievable TIN'' are represented by the first, the second
and the last line in (\ref{eq:sum slope 2-user equal power}), respectively.
The outer bound is given by (\ref{eq:coro2 1}). Given Corollary \ref{cor:low-snr noisy},
if $a\leq\frac{1}{4}$, treating interference as noise achieves optimal
the sum slope, i.e., the inner bound is tight, which is also indicated
on the figure.

The focal point here is the point $a=1$. Just above that, the effect
of interference is completely eliminated. Just below that, interference
is at its worst. One could wonder if, for the $K$-user case, the
former fact could be used effectively. It turns out that is not the
case. In Section \ref{Outer.sec}, we will show that in a $K$-user
interference channel when $K$ is large, with high probability each
user will form an 2-user weak interference pair, where $a$ is just
below 1, with some other user. 

\section{The $K$-user case}

\subsection{\label{sub: k-user achievable}Achievable Scheme and Inner Bound}

For the 2-user case, the achievable rates are unaffected by delay,
as seen in Section \ref{sub:2-user Achievable-Scheme}. However, the
$K$-user case is very different from the 2-user case, just as for
the high SNR case considered in \cite{CadambeJafar07}. Similar to
the high SNR case, we can obtain a significant increase in rate by
using a variation of interference alignment. The type of interference
alignment used in \cite{CadambeJafar07} based on time or frequency
selectivity does not work in the low-SNR regime; however, propagation
delays can be used. Specifically, we show that for any set of delays
$\tau_{ji}$, $i,j=1,\cdots,\, K$ that are linearly independent over
the rational numbers, there exist arbitrarily large $B$ that can
make the direct propagation delays $\tau_{ii}$ arbitrarily close
to an even integer while the cross-delays $\tau_{ji},j\neq i$ are
close to some odd integer. As a result, if we let each user use even
time slots in the discrete time baseband channel model, then at the
receiver, the desired message and the interference signal are almost
orthogonal in the time domain. Therefore, the interference channel
can achieve $\Delta\slope_{0}=\frac{1}{2}$.

The idea of interference alignment over time domain is also used in
\cite{CadJaf07ACSSC,GrokTseYat08}, and \cite{MaBo09}. However, delay
is much more efficient when we let $B\to\infty$, as $n_{ji}=\left\lfloor \tau_{ji}B\right\rfloor $
can become arbitrarily large. We therefore do not need to use the
approximation $\delta_{ji}\approx0$ as in \cite{CadJaf07ACSSC} or
large $K$ as in \cite{GrokTseYat08}. The paper \cite{MaBo09} mainly
discusses how to design an algorithm to place $K$ users in an $N$
dimensional space, $N>3$, so that interference alignment can be realized.
In our work, $N\geq2$ and user locations are given. Our method for
interference alignment works for any user location (with probability
1 for a continuous distribution of user locations).

In order to state the results, we need to refine the definition of
wideband slope (\ref{eq:slope}) as
\begin{eqnarray}
\ebno_{\min} & = & \liminf_{B\rightarrow+\infty}\frac{P}{R\cdot N_{0}B}\label{eq:ebno-1}\\
\mathcal{S}_{0} & \triangleq & \limsup_{\ebno\downarrow\ebno_{\min}}\frac{R\left(\ebno\right)}{10\log_{10}\ebno-10\log_{10}\ebno_{\min}}10\log_{10}2.\label{eq:slopegen}
\end{eqnarray}
Notice that if the limit exists, (\ref{eq:slopegen}) is identical
to (\ref{eq:slope}), so this is not a new definition, but a widening
of the applicability of the wideband slope. We will see through an
example in Section \ref{PracticalImplementation.sec} that this generalized
definition has an operational meaning, as does the wideband slope
in \cite{Ver02IT}.

For comparison purposes, we will list the results for the interference-free
case, i.e., $C_{ji}=0$ for $i\neq j$ as follows (directly obtained
from \cite[Theorem 9]{Ver02IT}):
\begin{eqnarray*}
\begin{array}{rcll}
\mathcal{S}_{0,\mathrm{no\, interference}} & = & 2{\displaystyle \frac{\left(\sum_{j}\left|C_{jj}\right|^{2}\right)^{2}}{\sum_{j}\left|C_{jj}\right|^{4}}} & \mbox{equal power\;\ constraint;}\\
\mathcal{S}_{0,\mathrm{no\, interference}} & = & 2K & \mbox{equal rate\;\ constraint}.
\end{array}
\end{eqnarray*}

\subsubsection{The Achievable Sum Slope}

In the following, we will precisely specify the interference alignment
scheme we use to obtain the achievable sum slope.
\begin{defn}[Delay-based interference alignment]
\label{def:The-transmission-scheme}Fix the transmission bandwidth
$B_{0}\leq B$.
\begin{enumerate}
\item At transmitter $j$

\begin{itemize}
\item Use a codebook generated from independent Gaussians according to 
\begin{equation}
x_{j}\left[n\right]\sim\left\{ \begin{array}{ll}
\mathcal{N}\left(0,\;\frac{2P_{j}}{B_{o}}\right) & n=2k\\
0 & n=2k+1
\end{array}\right.k\in\mathbb{Z}.\label{delaytrans.eq}
\end{equation}

\item Generate the baseband transmitted signal according to 
\[
\ensuremath{x_{j}\left(t\right)=\sum_{n=-\infty}^{+\infty}x_{j}\left[n\right]}\mathrm{sinc}\left(B_{o}\left(t-nT\right)\right).
\]
Notice that this signal has bandwidth $B_{0}\leq B$.
\end{itemize}
\item At receiver $j$: 

\begin{itemize}
\item Sample the received continuous time baseband signal with rate $B_{o}$
and symbol synchronize with $x_{j}[n]$;
\item Discard $y_{j}[2m+1],m\in\mathbb{Z}$
\item Decode the desired message from $y_{j}\left[2m\right],m\in\mathbb{Z}$,
with typical decoding while treating any remaining interference as
noise.
\end{itemize}
\end{enumerate}
\end{defn}
Since we only use every other time slot for transmission in (\ref{delaytrans.eq})
we can \emph{at most} achieve a wideband slope $\Delta S_{0}=\frac{1}{2}$.
In the following we will show that it is possible to choose the transmission
bandwidth $B_{0}$ so that most of the interference lines up in the
discarded time slots $y_{j}[2m+1],m\in\mathbb{Z}$, which in turn
means that we can actually achieve $\Delta S_{0}=\frac{1}{2}$.

The concept of\emph{ linearly independent over rational number} will
be introduced first.
\begin{defn}[\cite{Apostol-NumberTheory}]
A set of real numbers $\theta=\left\{ \theta_{1},\theta_{2},\cdots,\,\theta_{n}\right\} $
are linearly independent over rational number if $\sum a_{i}\theta_{i}=0$
only if $a_{i}=0$ for all $a_{i}\in\mathbb{Z}$.\end{defn}
\begin{lem}
\label{lem:approximation}If $\tau_{ji}$, $i,\, j\in\left\{ 1,\cdots,\, K\right\} ,i\neq j$
are linearly independent over the rational numbers, then for any $\delta>0$,
there exist an arbitrarily large real number $B$ , such that 
\begin{eqnarray}
\left|\tau_{ji}B-2k_{ji}-1\right| & \leq & \delta\label{eq:lemma dio}
\end{eqnarray}
for some integers $k_{ji}$, $j,\, i\in\left\{ 1,\cdots,\, K\right\} $. 
\end{lem}
The proof of Lemma \ref{lem:approximation} is based on the following
fundamental approximation results in number theory.
\begin{thm}[{\cite[Theorem 7.9, First Form of Kronecker's Theorem]{Apostol-NumberTheory}}]
\label{thm:Kronecker's}If $\alpha_{1},\cdots,\alpha_{n}$ are arbitrary
real numbers, if $\theta_{1},\cdots,\theta_{n}$ are linearly independent
real numbers over the rational numbers, and if $\epsilon>0$ is arbitrary,
then there exists an real number $t$ and integers $h_{1},\cdots,h_{n}$
such that 
\begin{eqnarray}
\left|t\theta_{i}-h_{i}-\alpha_{i}\right| & < & \epsilon\label{eq:kronecker's}
\end{eqnarray}
 $\forall i\in\left\{ 1,2,\ldots,\, n\right\} $
\end{thm}
We also have
\begin{lem}
\label{lem:Exercise-7.7}\cite[Exercise 7.7, page 160]{Apostol-NumberTheory}Under
the hypotheses of Theorem \ref{thm:Kronecker's}, if $T>0$ is given,
there exists a real number $t>T$ satisfying the $n$ inequalities
(\ref{eq:kronecker's}). 
\end{lem}
Now let us prove Lemma \ref{lem:approximation}. 
\begin{IEEEproof}[Proof of Lemma \ref{lem:approximation}]
Let $\alpha_{1},\cdots,\alpha_{n}=0.5$, $\epsilon=\frac{\delta}{2}$.
According to Theorem \ref{thm:Kronecker's}, there exist arbitrarily
large real number $\hat{B}$ and some integers $n_{ji}$, $i,\, j\in\left\{ 1,\cdots,\, K\right\} $
such that
\begin{eqnarray*}
\left|\tau_{ji}\hat{B}-k_{ji}-0.5\right| & \leq & \frac{\delta}{2}.
\end{eqnarray*}
Let $B=2\hat{B}$, we have 
\begin{eqnarray*}
\left|\tau_{ji}B-2k_{ji}-1\right| & \leq & \delta.
\end{eqnarray*}
Combining the inequality above with Lemma \ref{lem:Exercise-7.7},
Lemma \ref{lem:approximation} is proved.
\end{IEEEproof}
Lemma \ref{lem:approximation} shows that using this transmission
scheme, the desired signal is almost orthogonal with the interference
signal in time domain. However, there is always some interference
leaking into the signal time slots. We need to show that as the fractional
delay (\ref{fracdelay.eq}) $\delta_{ji}\to0$, the power of this
interference become negligible. For this we need the following lemma, 
\begin{lem}
\label{lem:mean square conv}Under the assumptions $x_{j}\left[2m\right]$
are i.i.d. Gaussian random variable with distribution $\mathcal{N}\left(0,\,2P_{j}\right)$
and $x_{j}\left[2m+1\right]=0$ for all $j$ and $m$, $\mathrm{E}\left[\tilde{x}_{i}[n_{1},\delta_{ji}]\tilde{x}_{i}^{*}[n_{2},\delta_{ji}]\right]$
is a continuous function of $\delta_{ji}$ which satisfies 
\begin{eqnarray*}
\lim_{\delta_{ji}\downarrow0}\mathrm{E}\left[\tilde{x}_{i}^{*}[n_{1},\delta_{ji}]\tilde{x}_{i}[n_{2},\delta_{ji}]\right] & = & \begin{cases}
2P_{i} & \mathrm{if}\; n_{1}=n_{2}=2k,\\
 & \mathrm{for\; some\; integer}\; k\\
0 & o.w.
\end{cases}
\end{eqnarray*}
\end{lem}
\begin{IEEEproof}
Please see Appendix \ref{sec:Proof-of-Lemma mean square}. 
\end{IEEEproof}
Equipped with the interference alignment scheme in definition \ref{def:The-transmission-scheme},
Lemma \ref{lem:approximation}, and Lemma \ref{lem:mean square conv},
we proceed to show the main results on the achievable sum slope of
the $K$-user interference channel.
\begin{thm}
\label{thm:intf alignment}Suppose that the set of delays $\tau_{ji},\, i,\, j\in\left\{ 1,\cdots,\, K\right\} ,i\neq j$
are linearly independent over the rational numbers. Then the following
wideband slope is achievable
\begin{eqnarray*}
\begin{array}{rcll}
\mathcal{S}_{0} & = & {\displaystyle \frac{\left(\sum_{j}\left|C_{jj}\right|^{2}\right)^{2}}{\sum_{j}\left|C_{jj}\right|^{4}}} & \mbox{equal power\;\ constraint;}\\
\mathcal{S}_{0} & = & K & \mbox{equal rate\;\ constraint}.
\end{array}
\end{eqnarray*}
Under both constraints,
\begin{eqnarray*}
\Delta\mathcal{S}_{0} & = & \frac{1}{2}
\end{eqnarray*}
is achievable.\end{thm}
\begin{IEEEproof}
Assume that the system uses the transmission scheme proposed in Definition
\ref{def:The-transmission-scheme}. Let
\begin{eqnarray*}
\epsilon_{j}(B) & = & \sum_{i\neq j}^{K}\left|\mathrm{E}\left[\left(\tilde{x}_{i}[2n]\right)^{2}\right]\right|
\end{eqnarray*}
denote the power of the leaked interference.

The best rate with this scheme is clearly achieved if the leaked interference
power is zero; in that case the channel is an interference-free channel
where half the symbols are not used. We can therefore conclude
\begin{eqnarray}
\Delta\mathcal{S}_{0} & \leq & \frac{1}{2}\label{achieveS0up.eq}
\end{eqnarray}

On the other hand, taking into account the leaked interference, the
achievable rate at receiver $j$ is 
\begin{eqnarray}
R_{j} & = & \frac{1}{2}\log\left(1+\frac{\left|C_{jj}\right|^{2}\frac{2P_{j}}{BN_{0}}}{1+\frac{\epsilon_{j}(B)}{BN_{0}}}\right)\label{eq:taylor 1}\\
 & = & \left|C_{jj}\right|^{2}\frac{P_{j}}{BN_{0}}-\left(\epsilon_{j}(B)\left|C_{jj}\right|^{2}P_{j}+\left|C_{jj}\right|^{4}P_{j}^{2}\right)\left(\frac{1}{BN_{0}}\right)^{2}+o\left(\left(\frac{1}{BN_{0}}\right)^{2}\right).\label{eq:taylor 2}
\end{eqnarray}
The wideband slope is a continuous function of the coefficients in
the first two terms in the Taylor series of $R_{j}$ in $\frac{1}{B}$.
According to Lemma \ref{lem:approximation} for any $\delta>0$ there
exists some $B_{\delta}$ and a set of integers $k_{ji}$ such that
$n_{ji}=2k_{ji}+1$, i.e., the integer part of the delay is an odd
number, and the fractional part of the delay satisfies $\left|\delta_{ji}\right|\leq\;\delta.$ 

From Lemma \ref{lem:approximation} and Lemma \ref{lem:mean square conv},
we can then conclude that there exists a sequence of real numbers
$\left\{ B_{o1},\, B_{o2},\cdots\right\} $, $B_{o\left(k+1\right)}>B_{ok}$,
so that $k\to\infty$ and $\epsilon_{j}(B_{ok})\to0$ for all $j=1,\cdots,\, K$.
This means that $\Delta\mathcal{S}_{0}=\frac{1}{2}$ is a limit point,
and together with (\ref{achieveS0up.eq}) this shows that $\Delta\mathcal{S}_{0}=\frac{1}{2}$
is the limit superior.
\end{IEEEproof}
The Theorem has the following corollary.
\begin{cor}
\label{cor:position}Suppose that all transmitters and receivers have
independent positions and each node position has a continuous distribution.
Then the propagation delays $\tau_{ji}$, $i,\, j\in\left\{ 1,\cdots,\, K\right\} $,
are linearly independent over the rational numbers with probability
one, and 
\begin{eqnarray*}
\Delta\mathcal{S}_{0} & = & \frac{1}{2}
\end{eqnarray*}
is achievable.
\end{cor}
So in practice $ $$\Delta\mathcal{S}_{0}=\frac{1}{2}$ is achievable,
since transmitters and receivers can never be positioned accurately
in a grid; there is always some nano-scale inaccuracy (dither) in
positions, at the fundamental level due to quantum mechanics!

\subsubsection{\label{PracticalImplementation.sec}Practical Implementation and
Simulation Results}

In this section we will show that the interference alignment ideas
of the previous section can be used in a practical system, and show
some simulation results. This will also make it clear why the modified
definition (\ref{eq:slopegen}) is needed.

We can see that one key question concerning the transmission scheme
defined by Definition \ref{def:The-transmission-scheme} is: how to
find $B_{o}$? Here we propose an algorithm, stated in the following
proposition.
\begin{prop}
\label{prop:find Bo}Assume that both the transmitters and the receivers
have perfect channel knowledge. \\
$\qquad$Initialize $B$ to be any positive integer. Proceed with
the following while loop:\\
$\qquad$While $\left(\exists i\neq j:\left|\tau_{ji}B-2k_{ji}-1\right|>\delta\right)$
$\{$\\
$\qquad$$\qquad$$\mathrm{Increase\;}B\;\mathrm{by}\;1,\, i.e.,\, B=B+1.$\\
$\qquad$$\}$

If $\tau_{ji}$ are linearly independent over the rational numbers
the algorithm terminates after a finite number of iterations. The
output $B$ of the algorithm satisfies 
\begin{eqnarray*}
\forall i\neq j:\left|\tau_{ji}B-2k_{ji}-1\right| & \leq & \delta,
\end{eqnarray*}
which can therefore can be chosen as $B_{o}$.
\end{prop}
Lemma \ref{lem:approximation-1} guarantees that the searching algorithm
defined in the proposition above terminates. The proof is almost identical
to the proof of Lemma \ref{lem:approximation}. However, the essential
difference is that while Lemma \ref{lem:approximation} only shows
the existence of $B$ satisfying (\ref{eq:lemma dio}) over the set
of positive real numbers $\mathbb{R}^{+}$, the results in this section
ensure that such $B$ can be found even if we restrict $B$ to be
integer. 
\begin{lem}
\label{lem:approximation-1}If $\tau_{ji}$ are linearly independent
over the rational numbers, then for any $\delta>0$, there exist an
integer $B$, such that 
\begin{eqnarray}
\left|\tau_{ji}B-2k_{ji}-1\right| & \leq & \delta\label{eq:lemma dio-1-1}
\end{eqnarray}
for some integers $k_{ji}$. Further, $B$ can be made arbitrarily
large. 
\end{lem}
The proof of Lemma \ref{lem:approximation-1} is based on the second
form of Kronecker's theorem\cite[Theorem 7.10, Second Form of Kronecker's Theorem]{Apostol-NumberTheory},
which shows that Theorem \ref{thm:Kronecker's} still holds even if
we require $t$ to be an integer. Details are skipped here.

\begin{figure}[tbh]
\subfloat[\label{simfig1.fig}Performance for arbitrary bandwidth.]{\includegraphics[width=3.5in]{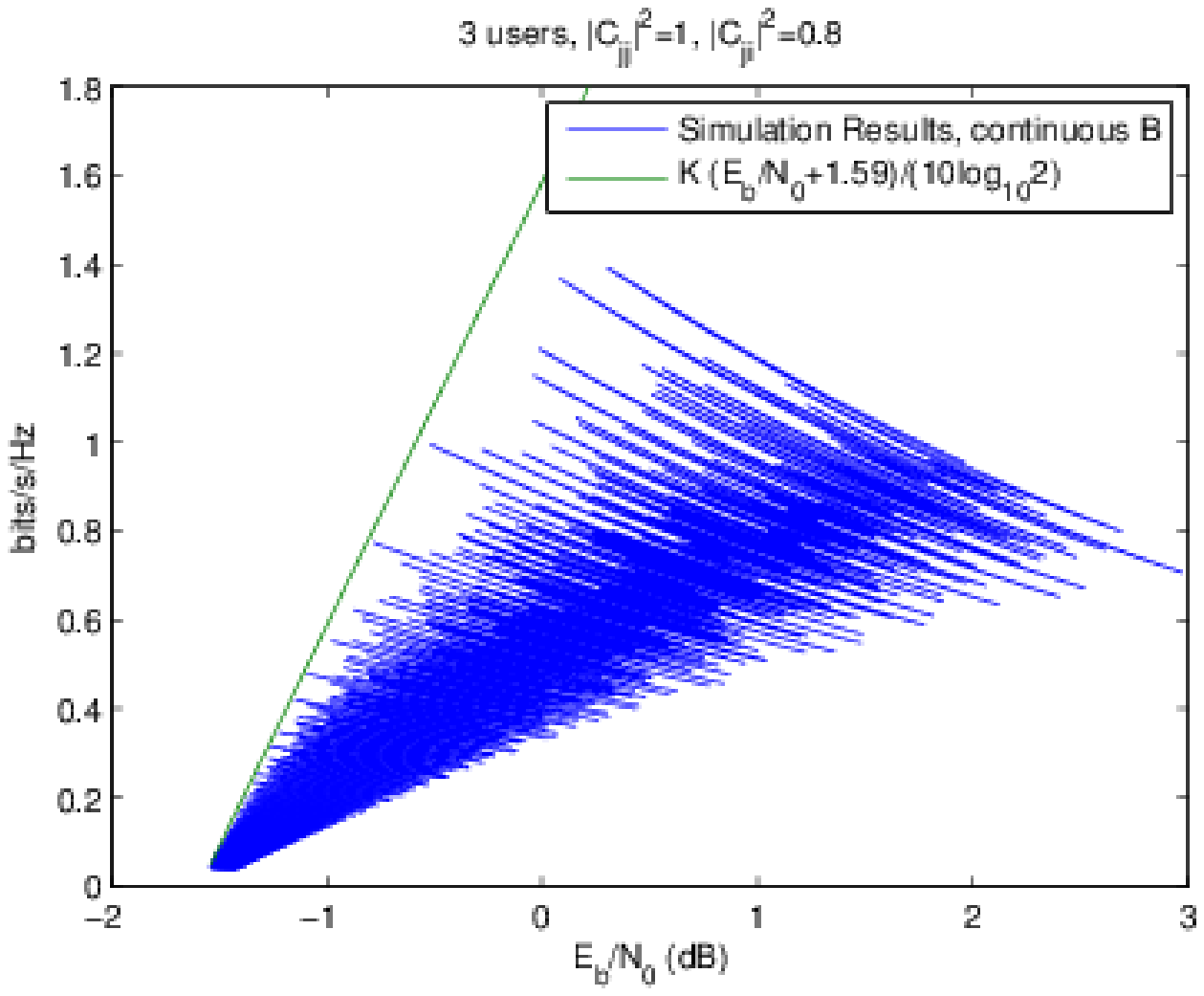}

}\subfloat[\label{simfig2.fig}Peak points of performance.]{\includegraphics[width=3.5in]{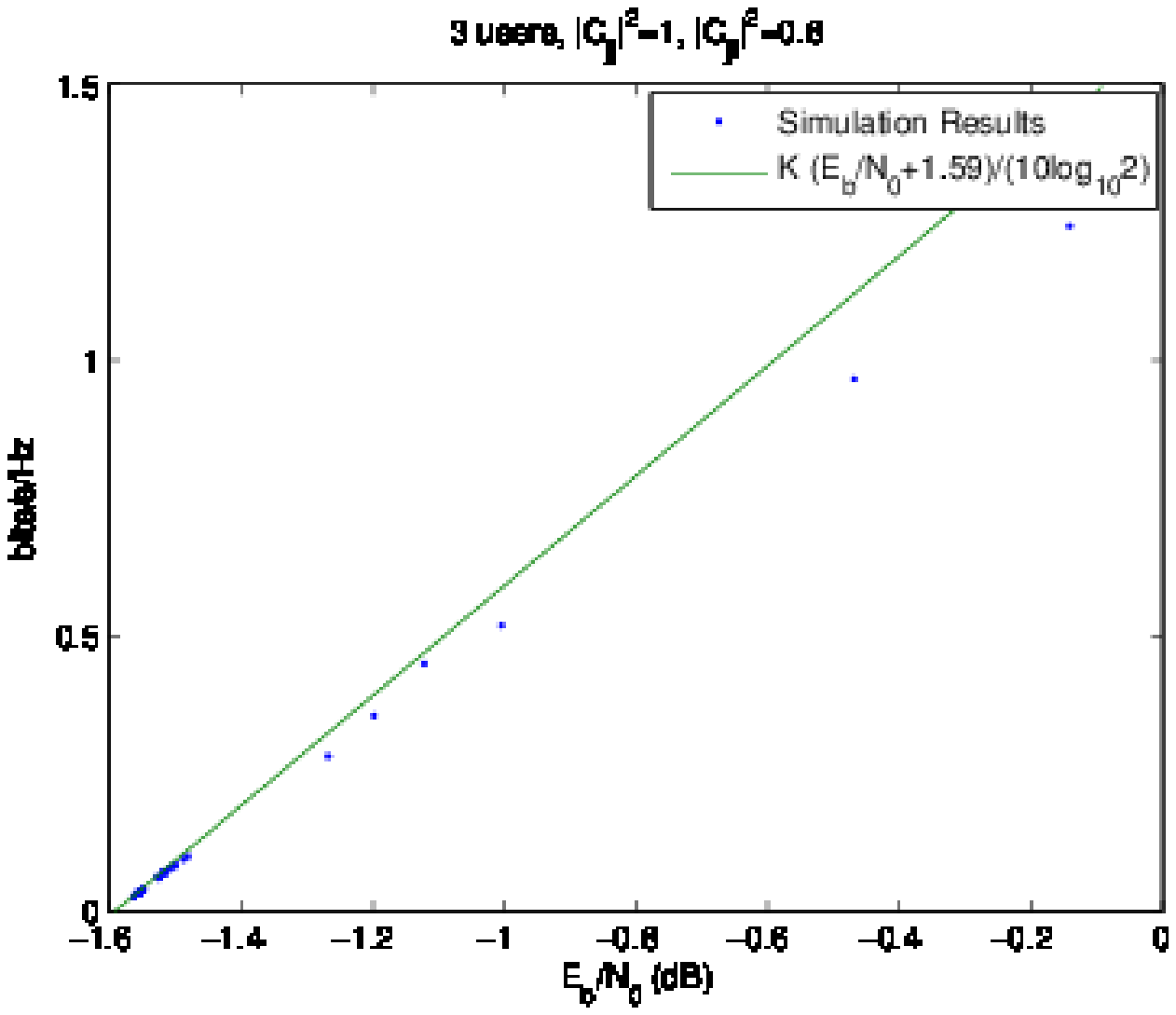}

}

\caption{Achievable spectral efficiency versus $\ebno$. The straight line
shows the performance approximated to first order by the wideband
slope.}
\end{figure}

We can see that the brute force algorithm of searching through all
integer $B$ is guaranteed to find good operating bandwidths. Fig.
\ref{simfig2.fig} shows the performance of the proposed achievable
scheme when the system operating at a sequence of $\mathrm{B}_{\delta}$,
$\delta=0.2$. However, designing more efficient $\mathrm{B}_{o}$-searching
algorithm could be a subject of further research. 

In the simulation, we consider a 3-user channel with symmetric channel
gain: $\left|C_{jj}\right|^{2}=1$, $\left|C_{ji}\right|^{2}=0.8$.
Notice that for channels with symmetric link gains, equal power and
equal rate constraints are equivalent. The delays $\tau_{ji}$ are
chosen such that they are linearly independent over the rational numbers. 

Fig. \ref{simfig1.fig} shows the simulation results of the case where
bandwidth $B$ increases continuously. The system performance shows
a noticeable oscillating behavior. This phenomenon can be explained
as followed. At receiver $j$ the interference caused by user $i$
is an increasing function of $\delta_{ji}$; and $\delta_{ji}$ is
a periodic function of $B$, oscillating between 0 and 1. It can be
proved that the cumulative effect of leaked interferences from all
other users has same (almost) periodic behavior. The proof is similar
to that of Lemmas \ref{OuterEqPower.eq} and \ref{lem:Exercise-7.7};
details are skipped here. 

Fig. \ref{simfig1.fig} also shows why we need the modified definition
(\ref{eq:slopegen}). In this case the limit (\ref{eq:slope}) does
not exist; one definition of a limit of a function is that for any
sequence $x_{n}\to x$, $f(x_{n})\to f(x)$. In Fig. \ref{simfig1.fig}
the points along the upper envelope and the points along the low envelope,
for example, give different slope. However, the $\limsup$ always
exists. One sequence that achieves the $\limsup$ is shown in Fig.
\ref{simfig2.fig}. What is important to notice that the new definition
of the wideband slope is still \emph{operational }as in \cite{Ver02IT}.
That is, it is possible to choose some finite bandwidth where the
performance is close to the wideband approximation. But different
from \cite{Ver02IT} is not enough to use a bandwidth that is sufficiently
large. It also has to be chosen very carefully.

\subsection{\label{Outer.sec}Outer bounds}

In Section \ref{sub: k-user achievable} we have seen that $\Delta\mathcal{S}_{0}=\frac{1}{2}$
can be achieved. Is this the best possible? Clearly no. In the 2-user
channel the interference alignment scheme proposed in Definition \ref{def:The-transmission-scheme}
reduces to TDMA. As we have seen in Section \ref{TwoUser.sec}, interference
decoding and treating interference as noise can be better than TDMA.
For $K>2$ case, it is also not difficult to construct examples where
$\Delta\mathcal{S}_{0}>\frac{1}{2}$ is achievable. However, in this
section we will show that for large $K$ this happens rarely. 

Let us first define two concepts: $\left(1-\epsilon\right)$-interference
pair and weak $\left(1-\epsilon\right)$-interference pair.
\begin{defn}
\label{def:1-e pair}We say that users $i$ and $j$ form an\emph{
$\left(1-\epsilon\right)$-interference pair} if
\[
1-\epsilon\leq\frac{\left|C_{ji}\right|^{2}}{\left|C_{ii}\right|^{2}},\frac{\left|C_{ij}\right|^{2}}{\left|C_{jj}\right|^{2}}<1,
\]
and form a \emph{weak $\left(1-\epsilon\right)$-interference} pair
if
\[
1-\epsilon\leq\frac{\left|C_{ji}\right|^{2}}{\left|C_{ii}\right|^{2}}<1\mbox{ or }1-\epsilon\leq\frac{\left|C_{ij}\right|^{2}}{\left|C_{jj}\right|^{2}}<1.
\]

\end{defn}
In section \ref{sub:outer Equal-Rate}, we will show that as the number
of users $K\rightarrow\infty$, the event 
\[
\left\{ \mbox{user\;}j,\forall j\in\left\{ 1,\cdots,K\right\} ,\mbox{\;\ forms\;\ a\;\ensuremath{\left(1-\epsilon\right)}-interference\;\ pair\;\ with\;\ at\;\ least\;\ one\;\ other\;\ user}\right\} 
\]
happens with high probability. Consider Fig. \ref{twouser.fig}: when
two users form an $(1-\epsilon)$-interference pair, they operate
in the point just below 1 in the figure, where Kramer's bound bounds
each user's wideband slope by 1. This results in $\Delta\mathcal{S}_{0}\leq\frac{1}{2}+\delta,\,\forall\delta>0$
under the equal rate constraint.

Similarly, in section \ref{sub:outer Equal-Power}, we will show that
as $K\rightarrow\infty$ , the event 
\[
\left\{ K\mbox{ users form }\frac{K}{2}\mbox{ disjoint weak }\left(1-\epsilon\right)-\mbox{interference pairs}\right\} 
\]
happens with high probability, which gives $\Delta\mathcal{S}_{0}\leq\frac{1}{2}+\delta,\,\forall\delta>0$
under the equal power constraint if the distribution of $C_{ji}$
satisfies some additional conditions. 

The outer bounds in this section are proven under the assumption that
the channel coefficients $C_{ji}$ for all $i,\, j\in\left\{ 1,\cdots,\, K\right\} $
are i.i.d. random variables. However, this is not a necessary condition,
only a convenient condition to simplify proofs; later in the section
we will comment more on this.

\subsubsection{\label{sub:outer Equal-Rate}The Equal Rate Constraint}

First consider the equal rate constraint. We assume that the channel
coefficients $C_{ji}$ are i.i.d. and $E\left[\left|C_{ii}\right|^{-2}\right]<\infty$;
if the latter assumption were not satisfied, $\lim_{K\to\infty}\frac{1}{K}\sum_{i=1}^{K}P_{i}=\infty$
even in the interference-free case (see (\ref{eq:ebnomin equal rate})),
so the energy per bit and wideband slope would not be well-defined
for large $K$ (see also the comment at the top of page 1325 in \cite{Ver02IT}
about Rayleigh fading). 

Let $F_{|C_{ii}|^{2}}$ be the CDF of $|C_{ij}|^{2}$; this defines
a probability measure on the real numbers through $\mu_{F}((a,b])=F_{|C_{ii}|^{2}}(b)-F_{|C_{ii}|^{2}}(b)$
(this is true for any random variable) For $\forall\epsilon,\,\hat{\epsilon}>0$,
define two sets 
\begin{eqnarray*}
R_{\hat{\epsilon}} & = & \{x\in\mathbb{R}:F_{\left|C_{ii}\right|^{2}}\left(x\right)-F_{\left|C_{ii}\right|^{2}}\left(\left(1-\epsilon\right)x\right)<\hat{\epsilon}\};\\
D_{\hat{\epsilon}} & = & \left\{ all\; i\in\{1,\ldots K\}:\left|C_{ii}\right|^{2}\in R_{\hat{\epsilon}}\right\} .
\end{eqnarray*}

The following lemma shows that as the number of users $K\rightarrow\infty$
a user in $D_{\hat{\epsilon}}^{c}$, with high probability forms a
$\left(1-\epsilon\right)$-interference pair with at least one other
user. 
\begin{lem}
\label{lem:F empty}Given $\forall\epsilon,\,\hat{\epsilon}>0$, denote
\begin{eqnarray*}
B_{\epsilon,\hat{\epsilon}} & = & \left\{ i\in D_{\hat{\epsilon}}^{c}:\mbox{user \ensuremath{i}does not form an \ensuremath{(1-\epsilon)}-inteference pair with any other user}\right\} 
\end{eqnarray*}
Then

\begin{eqnarray*}
\lim_{K\rightarrow\infty}\Pr\left(B_{\epsilon,\hat{\epsilon}}=\emptyset\right) & = & 1.
\end{eqnarray*}
\end{lem}
\begin{IEEEproof}
Please see Appendix \ref{sec:Proof-of-Lemma F empty}.
\end{IEEEproof}
On the other hand users in $D_{\hat{\epsilon}}$ do not necessarily
form $(1-\epsilon)$-pairs. The following lemmas are used to show
that the probability of the set $D_{\hat{\epsilon}}$ is small,
\begin{lem}
\label{lem:converge a}Given any infinite sequence $\hat{\epsilon}_{n}>0$
satisfying $\hat{\epsilon}_{n}>\hat{\epsilon}_{n+1}$ and $\hat{\epsilon}_{n}\rightarrow0$
, the corresponding sequence of $R_{\hat{\epsilon}_{n}}$ satisfies 

1) $R_{\hat{\epsilon}_{n+1}}\subseteq R_{\hat{\epsilon}_{n}}$;

2) $\mu_{F}(R_{\hat{\epsilon}_{n}})\rightarrow0$.\end{lem}
\begin{IEEEproof}
Please see Appendix \ref{sec:Proof-of-Lemma converge a}
\end{IEEEproof}

\begin{lem}
\label{lem:converge b} Let $X$ be a positive random variable with
$E[X]<\infty$, and let $\mu_{X}$ be the measure induced by the CDF
of $X$. Let $G_{i}\subset\mathbb{R}$ be a sequence of measurable
sets with $G_{i+1}\subseteq G_{i}$ and $\lim_{i\to\infty}\mu_{X}(G_{i})=0$.
Define 
\begin{eqnarray*}
X_{i} & = & \begin{cases}
X & X\in G_{i}\\
0 & X\notin G_{i}
\end{cases}.
\end{eqnarray*}
 Then
\begin{eqnarray*}
\lim_{i\to\infty}E[X_{i}] & = & 0.
\end{eqnarray*}
\end{lem}
\begin{IEEEproof}
Please see Appendix \ref{sec:Proof-of-Lemma converge b}.
\end{IEEEproof}
Our main result is stated in the following theorem.
\begin{thm}
\label{thm:eqrateup}Suppose that the channel coefficient $C_{ij}$
are i.i.d.. Under the equal rate constraint
\begin{eqnarray}
\forall\delta>0:\lim_{K\to\infty}\Pr\left(\Delta\mathcal{S}_{0}\leq\frac{1}{2}+\delta\right) & = & 1.\label{eqrateup.eq}
\end{eqnarray}
\end{thm}
\begin{IEEEproof}
We discuss users in the set $D_{\hat{\epsilon}}$ and those in the
set $D_{\hat{\epsilon}}^{c}$ separately. 

First, let us look at user $j$, $j\in D_{\hat{\epsilon}}^{c}$. We
assume that each user $j\in D_{\hat{\epsilon}}^{c}$ forms a $(1-\epsilon)$-interference
pair with some user $i_{(j)}$. Given Lemma \ref{lem:F empty}, this
happens with high probability. Consider a single $(1-\epsilon)$-interference
pair $(j,i_{(j)})$. We can get an upper bound on the spectral efficiency,
by eliminating all interference links except the links between users
$j$ and $i_{(j)}$, so that the received signal is
\begin{eqnarray*}
y_{j} & = & C_{jj}x_{j}+C_{ji_{\left(j\right)}}x_{i_{\left(j\right)}}+z_{j}\\
y_{i_{\left(j\right)}} & = & C_{i_{\left(j\right)}j}x_{j}+C_{i_{\left(j\right)}i_{\left(j\right)}}x_{i_{\left(j\right)}}+z_{i_{\left(j\right)}}.
\end{eqnarray*}
Let $\frac{\left|C_{ji_{\left(j\right)}}\right|^{2}}{\left|C_{i_{\left(j\right)}i_{\left(j\right)}}\right|^{2}}=1-\epsilon_{ji_{\left(j\right)}},\frac{\left|C_{i_{\left(j\right)}j}\right|^{2}}{\left|C_{jj}\right|^{2}}=1-\epsilon_{i_{\left(j\right)}j}$
. Since $\left\{ y_{j},y_{i_{\left(j\right)}}\right\} $ is a $\left(1-\epsilon\right)$-interference
pair, we have 
\begin{eqnarray}
0 & \leq & \epsilon_{ji_{\left(j\right)}},\,\epsilon_{i_{\left(j\right)}j}<\epsilon.\label{eq:epsilon bound}
\end{eqnarray}

Applying (\ref{2user eqrate 3.eq}) and (\ref{2user eqrate 4.eq})
to $\left\{ y_{j},y_{i_{\left(j\right)}}\right\} $, we have the optimum
solution 
\begin{eqnarray}
\snr_{i_{\left(j\right)}o} & = & \left|C_{i_{\left(j\right)}i_{\left(j\right)}}\right|^{-2}\cdot\frac{2^{\frac{R_{s}}{K}}\left(2^{\frac{R_{s}}{K}}-1\right)\left(1-\epsilon_{i_{\left(j\right)}j}\right)\epsilon_{ji_{\left(j\right)}}+\epsilon_{i_{\left(j\right)}j}\left(2^{\frac{R_{s}}{K}}-1\right)}{1-\left(1-\epsilon_{ji_{\left(j\right)}}\right)\left(1-\epsilon_{i_{\left(j\right)}j}\right)}\label{eq:pij}\\
\snr_{jo} & = & \left|C_{jj}\right|^{-2}\cdot\frac{2^{\frac{R_{s}}{K}}\left(2^{\frac{R_{s}}{K}}-1\right)\left(1-\epsilon_{ji_{\left(j\right)}}\right)\epsilon_{i_{\left(j\right)}j}+\epsilon_{ji_{\left(j\right)}}\left(2^{\frac{R_{s}}{K}}-1\right)}{1-\left(1-\epsilon_{ji_{\left(j\right)}}\right)\left(1-\epsilon_{i_{\left(j\right)}j}\right)}.\label{eq:pj-1}
\end{eqnarray}
And $\snr_{i_{\left(j\right)}}+\snr_{j}\geq\snr_{i_{\left(j\right)}o}+\snr_{jo}$.
Notice that the RHS of (\ref{eq:pij b}) and (\ref{eq:pj-1 b}) are
monotonically decreasing function of either $\epsilon_{ji_{\left(j\right)}}$
or $\epsilon_{i_{\left(j\right)}j}$. Thus, given the condition (\ref{eq:epsilon bound}),
we can relax (\ref{eq:pij b}) and (\ref{eq:pj-1 b}) by substituting
$\epsilon_{ji_{\left(j\right)}}$ and $\epsilon_{i_{\left(j\right)}j}$
by $\epsilon$,
\begin{eqnarray}
\snr_{i_{\left(j\right)}o} & \geq & \left|C_{i_{\left(j\right)}i_{\left(j\right)}}\right|^{-2}\cdot\frac{2^{\frac{R_{s}}{K}}\left(\left(1-\epsilon\right)2^{\frac{R_{s}}{K}}+\epsilon\right)-1}{2-\epsilon}\label{eq:pij b}\\
\snr_{jo} & \geq & \left|C_{jj}\right|^{-2}\cdot\frac{2^{\frac{R_{s}}{K}}\left(\left(1-\epsilon\right)2^{\frac{R_{s}}{K}}+\epsilon\right)-1}{2-\epsilon}.\label{eq:pj-1 b}
\end{eqnarray}
Thus 
\begin{eqnarray}
\snr_{jo} & = & \frac{2^{\frac{R_{s}}{K}}\left(\left(1-\epsilon\right)2^{\frac{R_{s}}{K}}+\epsilon\right)}{2-\epsilon}\left|C_{jj}\right|^{-2},\mathrm{if}\; j\in D_{\hat{\epsilon}}^{c}.\label{eq:jo}
\end{eqnarray}
 Second, for user $k$, $k\in D_{\hat{\epsilon}}$, we treat them
as being interference-free. In this case, we have
\begin{eqnarray}
\snr_{k} & \geq & \left(2^{\frac{R_{s}}{K}}-1\right)\left|C_{kk}\right|^{-2},\mathrm{if}\; k\in D_{\hat{\epsilon}}.\label{eq:ko}
\end{eqnarray}
Combining (\ref{eq:pj-1 b}) and (\ref{eq:ko}), the minimum sum power
required for an equal rate system with sum spectral efficiency $R_{s}$
is lower bounded by
\begin{eqnarray}
\snr_{s} & \geq & \frac{2^{\frac{R_{s}}{K}}\left(\left(1-\epsilon\right)2^{\frac{R_{s}}{K}}+\epsilon\right)}{2-\epsilon}\sum_{j\in D_{\hat{\epsilon}}^{c}}\left|C_{jj}\right|^{-2}\nonumber \\
 &  & +\left(2^{\frac{R_{s}}{K}}-1\right)\sum_{k\in D_{\hat{\epsilon}}}\left|C_{kk}\right|^{-2}.\label{eq:ps}
\end{eqnarray}
Using (\ref{eq:alt2}) on (\ref{eq:ps}) we get 
\begin{eqnarray}
\frac{E_{b}}{N_{0}}_{\mathrm{min}} & = & \frac{\sum\left(\left|C_{jj}\right|^{-2}\right)}{K}\log2\\
\Delta\mathcal{S}_{0} & = & \frac{\left(2-\epsilon\right)}{\left(4-3\epsilon\right)\left(1-\theta\right)+\left(2-\epsilon\right)\theta}\label{DeltaSproof.eq}
\end{eqnarray}
where 
\begin{eqnarray*}
\theta & \triangleq & \frac{\sum_{k\in D_{\hat{\epsilon}}}\left|C_{kk}\right|^{-2}}{\sum_{j=1}^{K}\left|C_{jj}\right|^{-2}}\;=\;\frac{\frac{1}{K}\sum_{k\in D_{\hat{\epsilon}}}\left|C_{kk}\right|^{-2}}{\frac{1}{K}\sum_{j=1}^{K}\left|C_{jj}\right|^{-2}}.
\end{eqnarray*}
Notice that the outer bound converges to the correct $\ebnomin$,
and (\ref{DeltaSproof.eq}) can therefore be used as an outer bound
on the slope.

Now, we want to show that $\forall\epsilon>0$, $\theta$ can be made
arbitrarily small. Define random variable $H_{j,\hat{\epsilon}}$
as
\begin{eqnarray*}
H_{j,\hat{\epsilon}} & = & \begin{cases}
\left|C_{jj}\right|^{-2} & j\in D_{\hat{\epsilon}}\\
0 & j\notin D_{\hat{\epsilon}}
\end{cases}.
\end{eqnarray*}
Given the fact that $H_{j,\hat{\epsilon}}$ and $H_{i,\hat{\epsilon}}$,
$i\neq j$ are independent, and $\sum_{k\in D_{\hat{\epsilon}}}\left|C_{kk}\right|^{-2}=\sum_{j=1}^{K}H_{j,\hat{\epsilon}}$,
we can apply the law of large number to $\theta$, which gives
\begin{eqnarray}
P\left(\lim_{K\rightarrow\infty}\theta=\frac{E\left(H_{j,\hat{\epsilon}}\right)}{E\left(\left|C_{jj}\right|^{-2}\right)}\right) & = & 1.\label{eq:theta1}
\end{eqnarray}
Combining Lemma \ref{lem:converge a} and Lemma \ref{lem:converge b},
we have 
\begin{eqnarray}
\lim_{\hat{\epsilon}\downarrow0}E\left(H_{j,\hat{\epsilon}}\right) & = & 0.\label{eq:theta2}
\end{eqnarray}

This proves (\ref{eqrateup.eq}) explicitly as follows. For any $\delta>0$
we can choose $\epsilon,\theta>0$ sufficiently small to make (\ref{DeltaSproof.eq})
less than $\frac{1}{2}+\delta$. We can choose $\hat{\epsilon}>0$
sufficiently small to make $\frac{E\left(H_{j,\hat{\epsilon}}\right)}{E\left(\left|C_{jj}\right|^{-2}\right)}$
smaller than $\theta$. Finally we can choose $K$ large enough to
make $\frac{\sum_{k\in D_{\hat{\epsilon}}}\left|C_{kk}\right|^{-2}}{\sum_{j=1}^{K}\left|C_{jj}\right|^{-2}}$
smaller than $\theta$ with high probability and $\Pr\left(B_{\epsilon,\hat{\epsilon}}=\emptyset\right)$
close to 1.
\end{IEEEproof}

\subsubsection{\label{sub:outer Equal-Power}The Equal Power Constraint}

We now consider the equal power constraint. Assume that the number
of users $K$ is an even integer, $K=2M$. For $\forall\epsilon>0$,
we define the event 
\[
A_{\epsilon}\triangleq\left\{ K\;\mathrm{users\; can\; form}\; M\; disjoint\;\mathrm{weak}\;(1-\epsilon)-\mathrm{pairs}\right\} ,
\]
and denote the indices of users belonging to the same weak $(1-\epsilon)$-pairs
as $\left\{ m_{1},\, m_{2}\right\} $.

Let the channel coefficients $C_{ij},i,j=1,\ldots,K$ be random variables
with a distribution that could depend on $K$. We consider the following
property of this sequence of distributions

\begin{property} \label{lem:matching}$\forall\epsilon:\mathrm{Pr}\left(A_{\epsilon}\right)\rightarrow1$
as $K\rightarrow\infty$. \end{property}
\begin{prop}
If the channel gains $C_{ij}$ are i.i.d (independent of $K$) with
continuous distribution, Property \ref{lem:matching} is satisfied.\end{prop}
\begin{IEEEproof}
Please see Appendix \ref{sec:Proof-of-Property matching}.\end{IEEEproof}
\begin{thm}
\label{thm:Upper Equal power}If property \ref{lem:matching} is satisfied
and the \emph{direct} channel gains $C_{jj}$ are i.i.d with finite
4th order moments, then under the equal power constraint
\begin{eqnarray}
\forall\delta>0:\lim_{K\to\infty}\Pr\left(\Delta\mathcal{S}_{0}\leq\frac{1}{\frac{\left(E\left[\left|C_{jj}\right|^{2}\right]\right)^{2}}{E\left[\left|C_{jj}\right|^{4}\right]}+1}+\delta\right) & = & 1.\label{OuterEqPower.eq}
\end{eqnarray}
\end{thm}
\begin{IEEEproof}
For the equal power constraint where $\snr_{j}=\frac{\snr_{s}}{K}$,
if property \ref{lem:matching} is satisfied, then for $K=2M$ users,
$M$ \emph{disjoint} weak $(1-\epsilon)$-pairs $\left\{ m_{1},\, m_{2}\right\} ,m=1,\cdots,M$
can be formed with high probability, and we will assume this is the
case. Applying Kramer's bound Theorem \ref{thm:Kramer} on each pair,
we have
\begin{eqnarray}
R_{m_{1}}+R_{m_{2}} & \leq & \min\left(\log\left(1+\left|C_{m_{1}m_{1}}\right|^{2}\frac{\snr_{s}}{K}\right)+\log\left(1+\frac{\left|C_{m_{2}m_{2}}\right|^{2}\nicefrac{\snr_{s}}{K}}{1+\left|C_{m_{2}m_{1}}\right|^{2}\frac{\snr_{s}}{K}}\right),\right.\nonumber \\
 &  & \left.\log\left(1+\left|C_{m_{2}m_{2}}\right|^{2}\frac{\snr_{s}}{K}\right)+\log\left(1+\frac{\left|C_{m_{1}m_{1}}\right|^{2}\nicefrac{\snr_{s}}{K}}{1+\left|C_{m_{1}m_{2}}\right|^{2}\frac{\snr_{s}}{K}}\right)\right)\label{eq:pair sum}
\end{eqnarray}
in $\mathrm{nats/s}$. For each weak $\left(1-\epsilon\right)$-pair,
(\ref{eq:pair sum}) gives 
\begin{eqnarray*}
\left.\frac{d\left(R_{m_{1}}+R_{m_{2}}\right)}{dP_{s}}\right|_{P_{s}=0} & = & \frac{\left|C_{m_{1}m_{1}}\right|^{2}+\left|C_{m_{2}m_{2}}\right|^{2}}{K}\\
-\left.\frac{d^{2}\left(R_{m_{1}}+R_{m_{2}}\right)}{d\snr_{s}^{2}}\right|_{P_{s}=0} & \geq & \frac{\left|C_{m_{1}m_{1}}\right|^{2}+\left|C_{m_{2}m_{2}}\right|^{2}+2\min\left\{ \left|C_{m_{1}m_{2}}\right|^{2}\left|C_{m_{1}m_{1}}\right|^{2},\left|C_{m_{2}m_{1}}\right|^{2}\left|C_{m_{2}m_{2}}\right|^{2}\right\} }{K^{2}}\\
 & = & \geq\frac{\left|C_{m_{1}m_{1}}\right|^{2}+\left|C_{m_{2}m_{2}}\right|^{2}+2\min\left\{ \left(1-\epsilon_{1}\right)\left|C_{m_{2}m_{2}}\right|^{2}\left|C_{m_{1}m_{1}}\right|^{2},\left(1-\epsilon_{2}\right)\left|C_{m_{1}m_{1}}\right|^{2}\left|C_{m_{2}m_{2}}\right|^{2}\right\} }{K^{2}}\\
 & \geq & \frac{\left|C_{m_{1}m_{1}}\right|^{2}+\left|C_{m_{2}m_{2}}\right|^{2}+2\left(1-\epsilon\right)\left|C_{m_{1}m_{1}}\right|^{2}\left|C_{m_{2}m_{2}}\right|^{2}}{K^{2}}
\end{eqnarray*}
since the $M$ pairs are disjoint and using the linearity of derivatives,
we have
\begin{eqnarray*}
\left.\frac{dR_{s}}{d\snr_{s}}\right|_{P_{s}=0} & = & \sum_{m=1}^{M}\left.\frac{d\left(R_{m_{1}}+R_{m_{2}}\right)}{d\snr_{s}}\right|_{P_{s}=0}\\
 & = & \frac{\sum_{j=1}^{K}\left|C_{jj}\right|^{2}}{K}\\
-\left.\frac{d^{2}R_{s}}{d\snr_{s}^{2}}\right|_{P_{s}=0} & = & \sum_{m=1}^{M}\left(-\left.\frac{d^{2}\left(R_{m_{1}}+R_{m_{2}}\right)}{d\snr_{s}^{2}}\right|_{P_{s}=0}\right)\\
 & \geq & \frac{\sum_{m=1}^{M}\left(\left|C_{m_{1}m_{1}}\right|^{4}+\left|C_{m_{2}m_{2}}\right|^{4}+2\left(1-\epsilon\right)\left|C_{m_{1}m_{1}}\right|^{2}\left|C_{m_{2}m_{2}}\right|^{2}\right)}{K^{2}}
\end{eqnarray*}
therefore 
\begin{eqnarray}
\ebno_{\min} & = & \frac{K\log_{e}2}{\sum_{j=1}^{K}\left|C_{jj}\right|^{2}}\nonumber \\
\mathcal{S}_{0} & \leq & 2\frac{\left(\sum_{j=1}^{K}\left|C_{jj}\right|^{2}\right)^{2}}{\sum_{m=1}^{M}\left(\left|C_{m_{1}m_{1}}\right|^{4}+\left|C_{m_{2}m_{2}}\right|^{4}+2\left(1-\epsilon\right)\left|C_{m_{1}m_{1}}\right|^{2}\left|C_{m_{2}m_{2}}\right|^{2}\right)}\nonumber \\
 & = & 2K\frac{\left(\frac{1}{K}\sum_{j=1}^{K}\left|C_{jj}\right|^{2}\right)^{2}}{\frac{1}{K}\sum_{j=1}^{K}\left|C_{jj}\right|^{4}+\left(1-\epsilon\right)\frac{1}{M}\sum_{m=1}^{M}\left|C_{m_{1}m_{1}}\right|^{2}\left|C_{m_{2}m_{2}}\right|^{2}}.\label{eq:thm27 s}
\end{eqnarray}
Now
\begin{eqnarray*}
\frac{1}{K}\sum_{j=1}^{K}\left|C_{jj}\right|^{2} & \stackrel{P}{\to} & \mathrm{E}\left[\left|C_{jj}\right|^{2}\right]\\
\frac{1}{K}\sum_{j=1}^{K}\left|C_{jj}\right|^{4} & \stackrel{P}{\to} & \mathrm{E}\left[\left|C_{jj}\right|^{4}\right]\\
\frac{1}{M}\sum_{m=1}^{M}\left|C_{m_{1}m_{1}}\right|^{2}\left|C_{m_{2}m_{2}}\right|^{2} & \stackrel{P}{\to} & \mathrm{E}\left[\left|C_{jj}\right|^{2}\right]^{2}
\end{eqnarray*}
as $K\to\infty$, where $\stackrel{P}{\to}$ stands for convergence
in probability since all random variables are positive and the moments
are assumed to exist. Using standard rules for convergence of transformation,
we then obtain
\begin{eqnarray*}
\forall\epsilon>0:\lim_{K\to\infty}\Pr\left(\Delta\mathcal{S}_{0}\leq\frac{1}{\left(1-\epsilon\right)\frac{\left(E\left[\left|C_{jj}\right|^{2}\right]\right)^{2}}{E\left[\left|C_{jj}\right|^{4}\right]}+1}\right) & = & 1
\end{eqnarray*}
from (\ref{eq:thm27 s}). Equivalently, 
\begin{eqnarray*}
\forall\epsilon>0:\lim_{K\to\infty}\Pr\left(\Delta\mathcal{S}_{0}\leq\frac{1}{\frac{\left(E\left[\left|C_{jj}\right|^{2}\right]\right)^{2}}{E\left[\left|C_{jj}\right|^{4}\right]}+1}+\delta_{\epsilon}\right) & = & 1
\end{eqnarray*}
where 
\begin{eqnarray*}
\delta_{\epsilon} & = & \frac{\epsilon\frac{\left(E\left[\left|C_{jj}\right|^{2}\right]\right)^{2}}{E\left[\left|C_{jj}\right|^{4}\right]}}{\left(1-\epsilon\right)\left(\frac{\left(E\left[\left|C_{jj}\right|^{2}\right]\right)^{2}}{E\left[\left|C_{jj}\right|^{4}\right]}\right)^{2}+\left(2-\epsilon\right)\frac{\left(E\left[\left|C_{jj}\right|^{2}\right]\right)^{2}}{E\left[\left|C_{jj}\right|^{4}\right]}+1}
\end{eqnarray*}
Notice that $\forall\delta$, $\exists\epsilon>0$ such that $\delta_{\epsilon}<\delta$.
Therefore, we obtain (\ref{OuterEqPower.eq}). 
\end{IEEEproof}
It is perhaps illustrative to write (\ref{OuterEqPower.eq}) in terms
of central moments,
\begin{eqnarray}
\frac{1}{\frac{\left(E\left[\left|C_{jj}\right|^{2}\right]\right)^{2}}{E\left[\left|C_{jj}\right|^{4}\right]}+1} & = & \frac{1}{\frac{\mu^{4}-2\sigma^{2}\mu^{2}+\sigma^{4}}{\mu^{4}+6\mu^{2}\sigma^{2}+4\gamma_{1}\mu\sigma^{3}+\gamma_{2}\sigma^{4}}+1}\label{OuterEqCentral.eq}\\
\mu & = & E[X]\nonumber \\
\sigma^{2} & = & \var[X]\nonumber \\
\gamma_{1} & = & \frac{E[(X-\mu)^{3}]}{\sigma^{3}}\nonumber \\
\gamma_{2} & = & \frac{E[(X-\mu)^{4}]}{\sigma^{4}}.\nonumber 
\end{eqnarray}
It can be seen that for $\sigma\ll\mu$, i.e., a nearly constant distribution,
(\ref{OuterEqCentral.eq}) is close to $\frac{1}{2}$.

We will discuss some implications of these theorems. For the equal
rate constraint, (\ref{eqrateup.eq}) essentially states that the
wideband slope is bounded by $\frac{1}{2}$ of that of no interference
for large $K$. Since this is also achievable by Theorem \ref{thm:intf alignment},
this is indeed the wideband slope, and delay-based interference alignment
is optimum. The bound for the equal power constraint is slightly weaker,
but is still close to $\frac{1}{2}$ for some distributions. 

Theorem \ref{thm:eqrateup} and \ref{thm:Upper Equal power} have
been proven under an i.i.d. assumption on all channel coefficients.
This can seem restrictive and not that realistic in a line of sight
model. However, the i.i.d. assumption is not essential. In Theorem
\ref{thm:eqrateup} it is used to prove that every user has at least
one other user with which it forms an $(1-\epsilon)$-pair with high
probability. This might be true under many other model assumptions.
It is also used to invoke the law of large numbers, which has a wide
range of generalizations. In Theorem \ref{thm:Upper Equal power}
the i.i.d. assumption is used to prove that users form disjoint weak
$(1-\epsilon)$-pairs, and again for invoking the law of large numbers.

What can be concluded is that for small special examples it is possible
to find a better wideband slope by optimizing a combination of interference
alignment, interference decoding, and treating interference as noise.
However, it probably does not pay off to try to find a general algorithm
for optimizing wideband slope: comparing the achievable sum slope
given by Theorem \ref{thm:intf alignment} and the upper bounds provided
by Theorem \ref{thm:eqrateup} and \ref{thm:Upper Equal power}, we
can see that as the number of users $K$ grows large, the gap between
the upper bounds and the inner bounds achieved by the interference
alignment scheme defined by Definition \ref{def:The-transmission-scheme}
becomes arbitrarily small. Furthermore, finding such schemes are hard
based on our experimentation.

Another interesting observation is that the outer bounds do not depend
on delay, only on the channel gains. Thus, the outer bound depends
on the macroscopic location of transmitters and receivers (e.g., if
gain is proportional to $d_{ji}^{\alpha}$ for some $\alpha>0$),
while the inner bounds depend on the microscopic location (i.e., fractional
delay differences). This also means that the outer bounds apply to
general scalar interference channels, not only LOS channels. However,
for non-LOS channels better outer bounds can be proven, which is the
subject of a later paper (for initial results, see \cite{ShenAHM11Allerton})

\section{Conclusions}

In this paper we have shown that by using interference alignment with
delay differences, a wideband slope of half of the interference-free
case is achievable. We have also shown that, mostly, it is the best
achievable. What it means is that near single-user performance can
be achieved in the interference channel in the low-$\mathrm{SNR}$
regime. One surprising conclusion is that orthogonalizing interference
is (near) optimum in the low-power regime. It is not too surprising
that this is optimum in the high-SNR regime \cite{CadambeJafar07},
since that regime is interference limited. But since the low-power
regime is also noise-limited, one could have expected that orthogonalizing
interference is sub-optimum. That is indeed the case for a 2-user
channel. But for a $K$-user channel, orthogonalizing is near optimum,
as shown by Theorem \ref{thm:Upper Equal power}.

A number of questions remain open. What if the bandwidth remains fixed,
but the transmission rate approaches zero (e.g., in a sensor network)?
This case is more complicated, and will be covered in a later paper.
How can the delay based interference alignment be implemented in practical
systems? As we have seen in section \ref{PracticalImplementation.sec},
the achievable spectral efficiency is very dependent on choosing the
right symbol rate, so this touches on issues of channel knowledge
and estimation, and how to optimize symbol rate in a given spectral
efficiency region, as well as up to what spectral efficiencies the
wideband slope provides a good approximation.

\bibliographystyle{IEEEtran}
\bibliography{Coop03,Coop06,wbslope}

\begin{thebibliography}{10}
\providecommand{\url}[1]{#1}
\csname url@samestyle\endcsname
\providecommand{\newblock}{\relax}
\providecommand{\bibinfo}[2]{#2}
\providecommand{\BIBentrySTDinterwordspacing}{\spaceskip=0pt\relax}
\providecommand{\BIBentryALTinterwordstretchfactor}{4}
\providecommand{\BIBentryALTinterwordspacing}{\spaceskip=\fontdimen2\font plus
\BIBentryALTinterwordstretchfactor\fontdimen3\font minus
  \fontdimen4\font\relax}
\providecommand{\BIBforeignlanguage}[2]{{%
\expandafter\ifx\csname l@#1\endcsname\relax
\typeout{** WARNING: IEEEtran.bst: No hyphenation pattern has been}%
\typeout{** loaded for the language `#1'. Using the pattern for}%
\typeout{** the default language instead.}%
\else
\language=\csname l@#1\endcsname
\fi
#2}}
\providecommand{\BIBdecl}{\relax}
\BIBdecl

\bibitem{AnnVee08}
V.~Sreekanth~Annapureddy and V.~Veeravalli, ``Sum capacity of the gaussian
  interference channel in the low interference regime,'' in \emph{Information
  Theory and Applications Workshop, 2008}, 27 2008-Feb. 1 2008, pp. 422--427.

\bibitem{MotKha08IT}
A.~Motahari and A.~Khandani, ``Capacity bounds for the gaussian interference
  channel,'' \emph{Information Theory, IEEE Transactions on}, vol.~55, no.~2,
  pp. 620--643, Feb. 2009.

\bibitem{ShaKraChe07IT}
X.~{Shang}, G.~{Kramer}, and B.~{Chen}, ``"{A New Outer Bound and the
  Noisy-Interference Sum-Rate Capacity for Gaussian Interference Channels}",''
  \emph{ArXiv e-prints}, vol. 712, dec 2007.

\bibitem{CadambeJafar07}
V.~Cadambe and S.~Jafar, ``Interference alignment and degrees of freedom of the
  $k$-user interference channel,'' \emph{Information Theory, IEEE Transactions
  on}, vol.~54, no.~8, pp. 3425 --3441, aug. 2008.

\bibitem{MaddahAliAl06}
M.~Maddah-Ali, A.~Motahari, and A.~Khandani, ``Signaling over {MIMO}
  multiple-base systems: combination of multiple-access and broadcast
  schemes,'' in \emph{Information {T}heory, 2006. {ISIT} 2006. {P}roceedings.
  {I}nternational {S}ymposium on}, 2006.

\bibitem{GomCadJaf08}
K.~S. Gomadam, V.~R. Cadambe, and S.~A. Jafar, ``Approaching the capacity of
  wireless networks through distributed interference alignment,'' \emph{CoRR},
  vol. abs/0803.3816, 2008.

\bibitem{CadJafSha09IT}
V.~R. Cadambe, S.~A. Jafar, and S.~Shamai, ``Interference alignment on the
  deterministic channel and application to fully connected gaussian
  interference networks,'' \emph{IEEE Trans. Inf. Theor.}, vol.~55, no.~1, pp.
  269--274, 2009.

\bibitem{CadambeJafarAl09}
V.~Cadambe, S.~A. Jafar, and C.~Wang, ``Interference alignment with asymmetric
  complex signaling - settling the h{\o}st-madsen-nosratinia conjecture,''
  \emph{{IEEE} Transactions on Information Theory}, submitted.

\bibitem{MotGhaKha09CoRR}
A.~S. Motahari, S.~O. Gharan, and A.~K. Khandani, ``Real interference alignment
  with real numbers,'' \emph{CoRR}, vol. abs/0908.1208, 2009.

\bibitem{EtkOrd09IT}
R.~H. Etkin and E.~Ordentlich, ``The degrees-of-freedom of the k-user gaussian
  interference channel is discontinuous at rational channel coefficients,''
  \emph{IEEE Trans. Inf. Theor.}, vol.~55, no.~11, pp. 4932--4946, 2009.

\bibitem{GhaMotKha09CoRR}
A.~Ghasemi, A.~S. Motahari, and A.~K. Khandani, ``Interference alignment for
  the k user mimo interference channel,'' \emph{CoRR}, vol. abs/0909.4604,
  2009.

\bibitem{Verdu90}
S.~Verdu, ``On channel capacity per unit cost,'' \emph{{IEEE} {T}ransactions on
  {I}nformation {T}heory}, vol.~36, no.~5, pp. 1019--1030, Sep. 1990.

\bibitem{Ver02IT}
S.~Verd{\'u}, ``Spectral efficiency in the wideband regime,'' \emph{IEEE
  Transactions on Information Theory}, vol.~48, no.~6, pp. 1319--1343, 2002.

\bibitem{CaireVerduAl04}
G.~Caire, D.~Tuninetti, and S.~Verdu, ``Suboptimality of {TDMA} in the
  low-power regime,'' \emph{IEEE Transactions on Information Theory}, vol.~50,
  no.~4, pp. 608--620, April 2004.

\bibitem{WieseAl10}
M.~Wiese, F.~Knabe, J.~G. Klotz, and A.~Sezgin, ``The performance of qpsk in
  low-snr interference channels,'' in \emph{2010 International Symposium on
  Information Theory and its Applications (ISITA), Taichung, Taiwan}, October
  2010.

\bibitem{CadJaf07ACSSC}
V.~Cadambe and S.~Jafar, ``Degrees of freedom of wireless networks - what a
  difference delay makes,'' nov. 2007, pp. 133 --137.

\bibitem{GrokTseYat08}
L.~Grokop, D.~N.~C. Tse, and R.~D. Yates, ``Interference alignment for
  line-of-sight channels,'' \emph{CoRR}, vol. abs/0809.3035, 2008.

\bibitem{MaBo09}
R.~Mathar and G.~B{\"o}cherer, ``On spatial patterns of transmitter-receiver
  pairs that allow for interference alignment by delay,'' in \emph{3rd
  International Conference on Signal Processing and Communication Systems
  (ICSPCS 2009)}, Omaha, USA, Sep. 2009.

\bibitem{F6}
``Fractionated spacecraft,''
  http://en.wikipedia.org/wiki/Fractionated\_spacecraft.

\bibitem{CaiTunVer04IT}
G.~Caire, D.~Tuninetti, and S.~Verd{\'u}, ``Suboptimality of tdma in the
  low-power regime,'' \emph{IEEE Transactions on Information Theory}, vol.~50,
  no.~4, pp. 608--620, 2004.

\bibitem{HanKobayashi81}
T.~Iian and K.~Kobayashi, ``A new achievable rate region for the interference
  channel,'' \emph{IEEE Transactions on Information Theory}, vol. IT-27, no.~1,
  pp. 49--60, January 1981.

\bibitem{EtkinTseWang07}
R.~H. Etkin, D.~N.~C. Tse, and H.~Wang, ``Gaussian interference channel
  capacity to within one bit,'' \emph{{IEEE} {T}ransactions on {I}nformation
  {T}heory}, submitted.

\bibitem{Kra04IT}
\BIBentryALTinterwordspacing
G.~Kramer, ``Outer bounds on the capacity of gaussian interference channels,''
  \emph{Information Theory, IEEE Transactions on}, vol.~50, no.~3, pp.
  581--586, 2004. [Online]. Available:
  \url{http://ieeexplore.ieee.org/xpls/abs\_all.jsp?arnumber=1273673}
\BIBentrySTDinterwordspacing

\bibitem{Cos85IT}
\BIBentryALTinterwordspacing
M.~Costa, ``On the gaussian interference channel,'' \emph{Information Theory,
  IEEE Transactions on}, vol.~31, no.~5, pp. 607--615, 1985. [Online].
  Available:
  \url{http://ieeexplore.ieee.org/xpls/abs\_all.jsp?arnumber=1057085}
\BIBentrySTDinterwordspacing

\bibitem{Bergmans74}
P.~Bergmans, ``A simple converse for broadcast channels with additive white
  {G}aussian noise,'' \emph{IEEE Transactions on Information Theory}, vol.
  IT-20, no.~2, pp. 279--280, March 1974.

\bibitem{ahm05ISIT}
A.~Host-Madsen and A.~Nosratinia, ``The multiplexing gain of wireless
  networks,'' in \emph{Information Theory, 2005. ISIT 2005. Proceedings.
  International Symposium on}, sept. 2005, pp. 2065 --2069.

\bibitem{ShaCheKra09}
\BIBentryALTinterwordspacing
X.~Shang, B.~Chen, G.~Kramer, and H.~V. Poor, ``Capacity regions and sum-rate
  capacities of vector gaussian interference channels,'' 2009. [Online].
  Available: \url{http://www.citebase.org/abstract?id=oai:arXiv.org:0907.0472}
\BIBentrySTDinterwordspacing

\bibitem{Apostol-NumberTheory}
T.~M. Apostol, \emph{Modular Functions and Dirichlet Series in Number Theory},
  2nd~ed.\hskip 1em plus 0.5em minus 0.4em\relax New York: Springer-Verlag
  Inc., 1990, graduate Texts in Mathematics;41.

\bibitem{ShenAHM11Allerton}
M.~Shen and A.~Hø~andst Madsen, ``Wideband slope of interference channel:
  Finite bandwidth case,'' in \emph{Communication, Control, and Computing
  (Allerton), 2011 49th Annual Allerton Conference on}, 2011, allerton 2011.

\bibitem{ShaCheKra08}
X.~{Shang}, B.~{Chen}, G.~{Kramer}, and H.~V. {Poor}, ``"{On the Capacity of
  MIMO Interference Channels}",'' \emph{ArXiv e-prints}, jul 2008.

\bibitem{rudin-principles}
W.~Rudin, \emph{Principles of mathematical analysis}, 3rd~ed.\hskip 1em plus
  0.5em minus 0.4em\relax New York: McGraw-Hill Book Co., 1976, international
  Series in Pure and Applied Mathematics.

\bibitem{bollobas01}
B.~Bollobas, \emph{Random Graphs}, W.~Fulton, A.~Katok, F.~Kirwan, P.~Sarnak,
  B.~Simon, and B.~Totaro, Eds.\hskip 1em plus 0.5em minus 0.4em\relax
  Cambridge University Press, 2001.

\bibitem{Janson00randomgraphs}
S.~Janson, ``Random graphs.''\hskip 1em plus 0.5em minus 0.4em\relax Wiley,
  2000.

\end{thebibliography}

\appendices{

\section{\label{sec:Proof-of-Theorem noisy with delay}Proof of Theorem \ref{thm:noisy with delay}}

The following lemmas will be used in this proof.
\begin{lem}
\label{lem:worst noise}Let $\underline{X}^{n}$=$\left\{ X_{1},X_{2},\cdots,X_{n}\right\} $
be a sequence of random variables satisfying power constraint $\frac{1}{n}\sum_{i=1}^{n}\cov\left(X_{i}\right)\leq\snr$.
Let $\underline{X}_{G}^{n}$=$\left\{ X_{1G},X_{2G},\cdots,X_{nG}\right\} $
be a sequence of i.i.d. Gaussian random variable, $X_{G}\sim\mathcal{N}\left(0,\,\snr\right)$.
Let $\underline{Z}_{1}^{n}$ and $\underline{Z}_{2}^{n}$ be two sequence
of i.i.d. random variables with distributions $Z_{1}\sim\mathcal{N}\left(0,\,\sigma_{1}^{2}\right)$
and $Z_{2}\sim\mathcal{N}\left(0,\,\sigma_{2}^{2}\right)$. Then we
have the following inequality
\begin{eqnarray*}
 &  & h\left(\underline{X}^{n}+\underline{Z}_{1}^{n}\right)-h\left(\underline{X}^{n}+\underline{Z}_{1}^{n}+\underline{Z}_{2}^{n}\right)\\
 & \leq & h\left(X_{G}+Z_{1}\right)-h\left(X_{G}+Z_{1}+Z_{2}\right).
\end{eqnarray*}

\end{lem}

\begin{lem}
\label{lem:two observations}Let $\underline{X}^{n}=\left\{ X_{1},X_{2}\cdots,\, X_{n}\right\} $
and $\underline{Y}^{n}=\left\{ Y_{1},\, Y_{2}\cdots,\, Y_{n}\right\} $
be two sequence of random variables. Let $\hat{X}_{G}$, $\check{X}_{G}$
and $\hat{Y}_{G}$, $\check{Y}_{G}$ be random variables satisfying
\begin{equation}
\cov\left(\begin{array}{c}
\hat{X}_{G}\\
\hat{Y}_{G}
\end{array}\right)\leq\frac{1}{n}\sum_{i=1}^{n}\cov\left(\begin{array}{c}
X_{i}\\
Y_{i}
\end{array}\right)\leq\cov\left(\begin{array}{c}
\check{X}_{G}\\
\check{Y}_{G}
\end{array}\right).\label{eq:two obs 1}
\end{equation}
Then 
\begin{align*}
h\left(\underline{X}^{n}\right) & \leq nh\left(\hat{X}_{G}\right)\leq nh\left(\check{X}_{G}\right)\\
h\left(\left.\underline{Y}^{n}\right|\underline{X}^{n}\right) & \leq nh\left(\left.\hat{Y}_{G}\right|\hat{X}_{G}\right)\leq nh\left(\left.\check{Y}_{G}\right|\check{X}_{G}\right).
\end{align*}
\end{lem}
\begin{IEEEproof}
This is a special case of \cite[Lemma 2]{ShaCheKra09}. 
\end{IEEEproof}
For the delay-free case where $\tilde{X}_{i}\left[n-n_{ji}\right]=X_{i}\left[n\right]$,
this theorem is identical to the previous results in \cite{ShaCheKra08,AnnVee08,MotKha08IT},
and a later work \cite{ShaCheKra09}. Here we use similar technique
as the proof of Theorem 6 in \cite{ShaCheKra09} to show that this
results still hold for channel with non-zero delay.

Assume that the channel coefficients and input power constraints satisfy
(\ref{eq:noisy condition}). Provide side information $S_{1}^{n}$
and $S_{2}^{n}$ to receiver 1 and 2 respectively 
\begin{eqnarray*}
S_{1}^{n} & = & C_{21}X_{1}^{n}+W_{1}^{n}\\
S_{2}^{n} & = & C_{12}X_{2}^{n}+W_{2}^{n}
\end{eqnarray*}
where $W_{j}$ are zero mean i.i.d Gaussian noise. And the joint distribution
of $W_{j}$ and $Z_{j}$ is 
\begin{eqnarray*}
\left(\begin{array}{c}
Z_{j}\\
W_{j}
\end{array}\right) & \sim & \mathcal{N}\left(0,\left(\begin{array}{cc}
1 & \rho_{j}\\
\rho_{j}^{*} & \sigma_{j}^{2}
\end{array}\right)\right),
\end{eqnarray*}
$\rho_{j}$ and $\sigma_{j}^{2}$ satisfy (\ref{eq:noisy cond 1})
to (\ref{eq:noisy cond 4}). From Fano's inequality, we have
\begin{eqnarray}
 &  & n\left(R_{1}+R_{2}\right)\nonumber \\
 & \leq & I\left(X_{1}^{n};\, Y_{1}^{n}\right)+I\left(X_{2}^{n};\, Y_{2}^{n}\right)+o(n)\nonumber \\
 & \leq & I\left(X_{1}^{n};\, Y_{1}^{n},\, S_{1}^{n}\right)+I\left(X_{2}^{n};\, Y_{2}^{n},\, S_{2}^{n}\right)+o(n)\nonumber \\
 & \overset{(a)}{=} & h\left(S_{1}^{n}\right)-h\left(\left.S_{1}^{n}\right|X_{1}^{n}\right)+h\left(\left.Y_{1}^{n}\right|S_{1}^{n}\right)-h\left(\left.Y_{1}^{n}\right|S_{1}^{n},\, X_{1}^{n}\right)\nonumber \\
 &  & +h\left(S_{2}^{n}\right)-h\left(\left.S_{2}^{n}\right|X_{2}^{n}\right)+h\left(\left.Y_{2}^{n}\right|S_{2}^{n}\right)-h\left(\left.Y_{2}^{n}\right|S_{2}^{n},\, X_{2}^{n}\right)\\
 & \overset{(b)}{=} & h\left(C_{21}X_{1}^{n}+W_{1}\right)-h\left(W_{1}^{n}\right)-h\left(\left.C_{12}\tilde{X}_{2}^{n}+Z_{1}^{n}\right|W_{1}^{n}\right)\nonumber \\
 &  & +h\left(\left.C_{11}X_{1}^{n}+C_{12}\tilde{X}_{2}^{n}+Z_{1}^{n}\right|C_{21}X_{1}^{n}+W_{1}^{n}\right)\nonumber \\
 &  & +h\left(C_{12}X_{2}^{n}+W_{2}^{n}\right)-h\left(W_{2}^{n}\right)-h\left(\left.C_{21}\tilde{X}_{1}^{n}+Z_{2}^{n}\right|W_{2}^{n}\right)\nonumber \\
 &  & +h\left(\left.C_{21}\tilde{X}_{1}^{n}+C_{22}X_{2}^{n}+Z_{2}^{n}\right|C_{12}X_{2}^{n}+W_{2}^{n}\right)+o(n)\\
 & \overset{(c)}{\leq} & -nh\left(W_{1}\right)+h\left(C_{12}X_{2}^{n}+W_{2}^{n}\right)-h\left(\left.C_{12}X_{2}^{n}+Z_{1}^{n}\right|W_{1}^{n}\right)\nonumber \\
 &  & +h\left(\left.C_{11}X_{1}^{n}+C_{12}\tilde{X}_{2}^{n}+Z_{1}^{n}\right|C_{21}X_{1}^{n}+W_{1}^{n}\right)\nonumber \\
 &  & -nh\left(W_{2}\right)+h\left(C_{21}X_{1}^{n}+W_{1}\right)-h\left(\left.C_{21}X_{1}^{n}+Z_{2}^{n}\right|W_{2}^{n}\right)\nonumber \\
 &  & +h\left(\left.C_{21}\tilde{X}_{1}^{n}+C_{22}X_{2}^{n}+Z_{2}^{n}\right|C_{12}X_{2}^{n}+W_{2}^{n}\right)+o(n)\\
 & \overset{(d)}{\leq} & -nh\left(W_{1}\right)+nh\left(C_{12}X_{2G}+W_{2}\right)-nh\left(\left.C_{12}X_{2G}+Z_{1}\right|W_{1}\right)\nonumber \\
 &  & +h\left(\left.C_{11}X_{1}^{n}+C_{12}\tilde{X}_{2}^{n}+Z_{1}^{n}\right|C_{21}X_{1}^{n}+W_{1}^{n}\right)\nonumber \\
 &  & -nh\left(W_{2}\right)+nh\left(C_{21}X_{1G}+W_{1}\right)-nh\left(\left.C_{21}X_{1G}+Z_{2}\right|W_{2}\right)\nonumber \\
 &  & +h\left(\left.C_{21}\tilde{X}_{1}^{n}+C_{22}X_{2}^{n}+Z_{2}^{n}\right|C_{12}X_{2}^{n}+W_{2}^{n}\right)+o(n)\\
 & \overset{(f)}{\leq} & -nh\left(W_{1}\right)+nh\left(C_{12}X_{2G}+W_{2}\right)-nh\left(\left.C_{12}X_{2G}+Z_{1}\right|W_{1}\right)\nonumber \\
 &  & +nh\left(\left.C_{11}X_{1G}+C_{12}\tilde{X}_{2G}+Z_{1}\right|C_{21}X_{1G}+W_{1}\right)\nonumber \\
 &  & -nh\left(W_{2}\right)+nh\left(C_{21}X_{1G}+W_{1}\right)-nh\left(\left.C_{21}X_{1G}+Z_{2}\right|W_{2}\right)\nonumber \\
 &  & +nh\left(\left.C_{21}\tilde{X}_{1G}+C_{22}X_{2G}+Z_{2}\right|C_{12}X_{2G}+W_{2}\right)+o(n)\\
 & \overset{(g)}{\leq} & -nh\left(W_{1}\right)+nh\left(C_{12}X_{2G}+W_{2}\right)-nh\left(\left.C_{12}X_{2G}+Z_{1}\right|W_{1}\right)\nonumber \\
 &  & +nh\left(\left.C_{11}X_{1G}+C_{12}X_{2G}+Z_{1}\right|C_{21}X_{1G}+W_{1}\right)\nonumber \\
 &  & -nh\left(W_{2}\right)+nh\left(C_{21}X_{1G}+W_{1}\right)-nh\left(\left.C_{21}X_{1G}+Z_{2}\right|W_{2}\right)\nonumber \\
 &  & +nh\left(\left.C_{21}X_{1G}+C_{22}X_{2G}+Z_{2}\right|C_{12}X_{2G}+W_{2}\right)+o(n)
\end{eqnarray}
where $\lim_{n\to\infty}o(n)/n=0$, $X_{jG}$ with '$G$' subscription
means that input at transmitter $j$ is i.i.d. Gaussian, with distribution
$X_{jG}\sim\mathcal{N}\left(0,\,\snr_{j}\right)$. $\left(a\right)$
is from chain rule. $(c)$ holds because both $X_{j}$ and $\tilde{X}_{j}$
can be obtained from sampling the same continuous-time baseband signal
$X_{j}\left(t\right)$ at the Nyquist rate, while $Z_{j}$ and $W_{j}$
are sampled from white Gaussian noise, so that 
\begin{eqnarray*}
h\left(\left.C_{21}\tilde{X}_{1}^{n}+Z_{2}^{n}\right|W_{2}^{n}\right) & = & h\left(\left.C_{21}X_{1}^{n}+Z_{2}^{n}\right|W_{2}^{n}\right)+o(n)\\
h\left(\left.C_{12}\tilde{X}_{2}^{n}+Z_{1}^{n}\right|W_{1}^{n}\right) & = & h\left(\left.C_{12}X_{2}^{n}+Z_{1}^{n}\right|W_{1}^{n}\right)+o(n)
\end{eqnarray*}
because Given (\ref{eq:noisy cond 1}) and (\ref{eq:noisy cond 2}),
$\cov\left(W_{1}^{n}\right)\leq\cov\left(\left.Z_{2}^{n}\right|W_{2}^{n}\right)$
and $\cov\left(W_{2}^{n}\right)\leq\cov\left(\left.Z_{1}^{n}\right|W_{1}^{n}\right)$.
Combining Lemma \ref{lem:worst noise} and \cite[Lemma 3]{ShaKraChe07IT},
we have 
\begin{eqnarray}
 &  & h\left(C_{12}X_{2}^{n}+W_{2}^{n}\right)-h\left(\left.C_{12}X_{2}^{n}+Z_{1}^{n}\right|W_{1}^{n}\right)\nonumber \\
 & \leq & nh\left(C_{12}X_{2G}+W_{2}\right)-nh\left(\left.C_{12}X_{2G}+Z_{1}\right|W_{1}\right)\label{eq:d_1}
\end{eqnarray}
and 
\begin{eqnarray}
 &  & h\left(C_{21}X_{1}^{n}+W_{1}^{n}\right)-h\left(\left.C_{21}X_{1}^{n}+Z_{2}^{n}\right|W_{2}^{n}\right)\nonumber \\
 & \leq & nh\left(C_{21}X_{1G}+W_{1}\right)-nh\left(\left.C_{21}X_{1G}+Z_{2}\right|W_{2}\right).\label{eq:d_2}
\end{eqnarray}
Therefore (d) is true. 

(f) is from Lemma \ref{lem:two observations}, where $\tilde{X}_{jG}$
are i.i.d. Gaussian random variable satisfying $\cov\left(\tilde{X}_{jG}\right)=\frac{1}{n}\trace\left(\tilde{X}_{j}^{n}\left(\tilde{X}_{j}^{n}\right)^{\dagger}\right)$.
Denote $\snr_{j}^{\prime}\triangleq\frac{1}{n}\trace\left(\tilde{X}_{j}^{n}\left(\tilde{X}_{j}^{n}\right)^{\dagger}\right)$
as the power of $\tilde{X}_{j}^{n}$. We could see that $\snr_{j}^{\prime}\leq\snr_{j}$,
because time-shifting of a signal sampled at the Nyquist rate does
not change signal power. Therefore we have (g). This shows that the
sum capacity of a channel with delay is outer bounded by that of a
channel without delay.

We could see that the inequality (g) is independent of the propagation
delay. It is identical to the first inequality in \cite[(89)]{ShaCheKra09}.
Therefore, from this point on, the proof is the same as in the delay-free
case.

\section{\label{sec:Proof-of-Lemma mean square}Proof of Lemma \ref{lem:mean square conv}}
\begin{IEEEproof}
We have 
\begin{eqnarray}
\mathrm{E}\left[\tilde{x}_{i}^{*}[n_{1},\delta_{ji}]\tilde{x}_{i}[n_{2},\delta_{ji}]\right] & = & \mathrm{E}\left[\left(\sum_{m=-\infty}^{\infty}x_{i}[2m]\sinc(n_{1}-2m+\delta_{ji})\right)^{*}\cdot\right.\nonumber \\
 &  & \left.\left(\sum_{m=-\infty}^{\infty}x_{i}[2m]\sinc(n_{2}-2m+\delta_{ji})\right)\right]\nonumber \\
 & = & \sum_{m=-\infty}^{\infty}\mathrm{E}\left[\left|x_{i}[2m]\right|^{2}\right]\sinc(n_{1}-2m+\delta_{ji})\sinc(n_{2}-2m+\delta_{ji})\nonumber \\
 & = & \sum_{m=0}^{\infty}\mathrm{E}\left[\left|x_{i}[2m]\right|^{2}\right]\sinc(n_{1}-2m+\delta_{ji})\sinc(n_{2}-2m+\delta_{ji})\nonumber \\
 &  & +\sum_{m=1}^{\infty}\mathrm{E}\left[\left|x_{i}[-2m]\right|^{2}\right]\sinc(n_{1}+2m+\delta_{ji})\sinc(n_{2}+2m+\delta_{ji})\nonumber \\
 & =2P_{i} & \left(\sum_{m=0}^{\infty}\sinc(n_{1}-2m+\delta_{ji})\sinc(n_{2}-2m+\delta_{ji})\right.\nonumber \\
 &  & \left.+\sum_{m=1}^{\infty}\sinc(n_{1}+2m+\delta_{ji})\sinc(n_{2}+2m+\delta_{ji})\right).\label{eq:cov n1n2-1}
\end{eqnarray}
Define $f_{m}\left(\delta_{ji}\right)$ and $g_{m}\left(\delta_{ji}\right)$
as 
\begin{eqnarray*}
f_{m}\left(\delta_{ji}\right) & \triangleq & \sinc(n_{1}-2m+\delta_{ji})\sinc(n_{2}-2m+\delta_{ji})\\
g_{m}\left(\delta_{ji}\right) & \triangleq & \sinc(n_{1}+2m+\delta_{ji})\sinc(n_{2}+2m+\delta_{ji})
\end{eqnarray*}
and their partial sums $s_{f,M}\left(\delta_{ji}\right)=\sum_{m=0}^{M}f_{m}\left(\delta_{ji}\right)$,
$s_{f}\left(\delta_{ji}\right)\triangleq\lim_{M\rightarrow\infty}s_{f,M}\left(\delta_{ji}\right)$;
$s_{g,M}\left(\delta_{ji}\right)=\sum_{m=0}^{M}g_{m}\left(\delta_{ji}\right)$,
$s_{g}\left(\delta_{ji}\right)\triangleq\lim_{M\rightarrow\infty}s_{g,M}\left(\delta_{ji}\right)$.
Here 
\begin{eqnarray*}
\left|f_{m}\left(\delta_{ji}\right)\right| & = & \left|\frac{\sin\left(\pi(n_{1}-2m+\delta_{ji})\right)\sin\left(\pi(n_{2}-2m+\delta_{ji})\right)}{(n_{1}-2m+\delta_{ji})(n_{2}-2m+\delta_{ji})}\right|\\
 & \leq & \frac{1}{(n_{1}-2m+\delta_{ji})(n_{2}-2m+\delta_{ji})}
\end{eqnarray*}
and
\begin{eqnarray*}
\left|g_{m}\left(\delta_{ji}\right)\right| & \leq & \frac{1}{(n_{1}+2m+\delta_{ji})(n_{2}+2m+\delta_{ji})}.
\end{eqnarray*}
Let $M_{f,k}\triangleq\frac{1}{(n_{1}-2m+\delta_{ji})(n_{2}-2m+\delta_{ji})}$,
$M_{g,k}\triangleq\frac{1}{(n_{1}+2m+\delta_{ji})(n_{2}+2m+\delta_{ji})}$.
Because $\sum_{m=1}^{\infty}\frac{1}{k^{2}}$ is convergent, $\sum_{k=0}^{\infty}M_{f,k}$
and $\sum_{k=1}^{\infty}M_{g,k}$ converge too. Due to Weierstrass's
test for uniform convergence\cite{rudin-principles}, $s_{f,M}\left(\delta_{ji}\right)$
and $s_{g,M}\left(\delta_{ji}\right)$ converge uniformly. And using
Theorem 7.11 in \cite{rudin-principles}, we have 
\begin{eqnarray*}
\lim_{\delta_{ji}\downarrow0}\lim_{M\rightarrow\infty}s_{f,M}\left(\delta_{ji}\right) & = & \lim_{M\rightarrow\infty}\lim_{\delta_{ji}\downarrow0}s_{f,M}\left(\delta_{ji}\right)\\
\lim_{\delta_{ji}\downarrow0}\lim_{M\rightarrow\infty}s_{g,M}\left(\delta_{ji}\right) & = & \lim_{M\rightarrow\infty}\lim_{\delta_{ji}\downarrow0}s_{g,M}\left(\delta_{ji}\right).
\end{eqnarray*}
Thus, (\ref{eq:cov n1n2-1}) becomes 
\begin{eqnarray*}
\lim_{\delta_{ji}\downarrow0}\mathrm{E}\left[\tilde{x}_{i}^{*}[n_{1},\delta_{ji}]\tilde{x}_{i}[n_{2},\delta_{ji}]\right] & = & 2P_{i}\left(\lim_{M\rightarrow\infty}\lim_{\delta_{ji}\downarrow0}s_{f,M}\left(\delta_{ji}\right)\right.\\
 &  & +\left.\lim_{M\rightarrow\infty}\lim_{\delta_{ji}\downarrow0}s_{g,M}\left(\delta_{ji}\right)\right)\\
 & = & \begin{cases}
2P_{i} & \mathrm{if}\; n_{1}=n_{2}=2k,\\
 & \mathrm{for\; some\; integer}\; k\\
0 & o.w.
\end{cases}
\end{eqnarray*}
And given Theorem 7.12 in \cite{rudin-principles} and the continuity
of $\sinc$ function, we can conclude that $\mathrm{E}\left[\tilde{x}_{i}^{*}[n_{1},\delta_{ji}]\tilde{x}_{i}[n_{2},\delta_{ji}]\right]$is
a continuous function of $\delta_{ji}$. 
\end{IEEEproof}

\section{\label{sec:Proof-of-Lemma F empty}Proof of Lemma \ref{lem:F empty}}

Let $C_{i,\epsilon}$ be the event that user$i$ does not form an
$\left(1-\epsilon\right)$-pair with any other user;. Then
\begin{eqnarray*}
\Pr(B_{\epsilon,\hat{\epsilon}}\neq\emptyset) & = & \Pr\left(\bigcup_{i=1}^{K}C_{i,\epsilon}\right)\\
 & \leq & \sum_{i=1}^{K}\Pr(C_{i,\epsilon})\\
 & = & K\Pr(C_{1,\epsilon})
\end{eqnarray*}
and 
\begin{eqnarray*}
\Pr(C_{1,\epsilon}) & = & \mathrm{Pr}\left(\forall j>1:\;\frac{\left|C_{j1}\right|^{2}}{\left|C_{11}\right|^{2}}or\frac{\left|C_{1j}\right|^{2}}{\left|C_{jj}\right|^{2}}\notin\left(1-\epsilon,1\right)\mbox{ and }\left|C_{11}\right|^{2}\notin R_{\hat{\epsilon}}\right)\\
 & = & \mathrm{Pr}\left(\forall j>1:\;\frac{\left|C_{j1}\right|^{2}}{\left|C_{11}\right|^{2}}\notin\left(1-\epsilon,1\right)\mbox{ and }\left|C_{11}\right|^{2}\notin R_{\hat{\epsilon}}\right)+\mathrm{Pr}\left(\forall j>1:\frac{\left|C_{1j}\right|^{2}}{\left|C_{jj}\right|^{2}}\notin\left(1-\epsilon,1\right)\right)\\
 &  & -\mathrm{Pr}\left(\forall j>1:\;\frac{\left|C_{j1}\right|^{2}}{\left|C_{11}\right|^{2}}\mbox{ and }\frac{\left|C_{1j}\right|^{2}}{\left|C_{jj}\right|^{2}}\notin\left(1-\epsilon,1\right)\mbox{ and }\left|C_{11}\right|\notin R_{\hat{\epsilon}}\right)\\
 & = & \left(1-p_{j}^{K-1}\right)\mathrm{Pr}\left(\forall j>1:\;\frac{\left|C_{j1}\right|^{2}}{\left|C_{11}\right|^{2}}\notin\left(1-\epsilon,1\right)\mathbf{\mbox{ and }}\left|C_{11}\right|^{2}\notin R_{\hat{\epsilon}}\right)+p_{j}^{K-1}
\end{eqnarray*}
where $p_{j}=\mathrm{Pr}\left(\frac{\left|C_{1j}\right|^{2}}{\left|C_{jj}\right|^{2}}\notin\left(1-\epsilon,1\right)\right)\in\left(0,1\right)$.
Notice that the events $\frac{\left|C_{1j}\right|^{2}}{\left|C_{jj}\right|^{2}}\notin\left(1-\epsilon,1\right)$
are independent for different $j$, and $p_{j}^{K-1}\rightarrow0$.
Thus, $\Pr(C_{1,\epsilon})\rightarrow\mathrm{Pr}\left(\forall j>1:\;\frac{\left|C_{j1}\right|^{2}}{\left|C_{11}\right|^{2}}\notin\left(1-\epsilon,1\right)\mbox{ and }\left|C_{11}\right|^{2}\notin R_{\hat{\epsilon}}\right)$.
As the $\left|C_{jj}\right|^{2}$ are independent, 
\begin{eqnarray*}
 &  & \mathrm{Pr}\left(\forall j>1:\;\frac{\left|C_{j1}\right|^{2}}{\left|C_{11}\right|^{2}}\notin\left(1-\epsilon,1\right)\mbox{ and }\left|C_{11}\right|^{2}\notin R_{\hat{\epsilon}}\right)\\
 & = & \int_{x\notin R_{\hat{\epsilon}}}\prod_{j=2}^{K}\left(1-\int_{\left(1-\epsilon\right)x}^{x}dF_{\left|C_{j1}\right|^{2}}(u)\right)dF_{\left|C_{11}\right|^{2}}(x)\\
 & = & \int_{x\notin R_{\hat{\epsilon}}}\left(1-F_{\left|C_{ii}\right|^{2}}\left(x\right)+F_{\left|C_{ii}\right|^{2}}\left(\left(1-\epsilon\right)x\right)\right)^{K-1}dF_{\left|C_{ii}\right|^{2}}\left(x\right)\\
 & \leq & \left(1-\hat{\epsilon}\right)^{K-1}\int_{x\notin R_{\hat{\epsilon}}}dF_{\left|C_{ii}\right|^{2}}\left(x\right)\\
 & = & \left(1-\mu_{F}(R_{\hat{\epsilon}})\right)\left(1-\hat{\epsilon}\right)^{K-1}.
\end{eqnarray*}
Thus, $P_{\sigma}\leq K\left(1-\mu_{F}(R_{\hat{\epsilon}})\right)\left(1-\hat{\epsilon}\right)^{K-1}$,
and $\lim_{K\rightarrow\infty}\Pr(B_{\epsilon,\hat{\epsilon}}\neq\emptyset)=0$.

\section{\label{sec:Proof-of-Lemma converge a}Proof of Lemma \ref{lem:converge a}}

Given the definition $\forall x\in R_{\hat{\epsilon}}:F_{\left|C_{ii}\right|^{2}}\left(x\right)-F_{\left|C_{ii}\right|^{2}}\left(\left(1-\epsilon\right)x\right)\leq\hat{\epsilon}$,
and given the fact $F_{\left|C_{ii}\right|^{2}}\left(x\right)-F_{\left|C_{ii}\right|^{2}}\left(\left(1-\epsilon\right)x\right)<\hat{\epsilon}_{n+1}<\hat{\epsilon}_{n}$,
it clearly follows that $R_{\hat{\epsilon}_{n+1}}\subseteq R_{\hat{\epsilon}_{n}}$.
Let $I_{\hat{\epsilon}_{n}}(x)$ be the indicator function of $R_{\hat{\epsilon}_{n}}$.
Using Lebesgue dominated convergence we have 
\begin{eqnarray*}
 &  & \lim_{n\rightarrow\infty}\int_{0}^{\infty}I_{\hat{\epsilon}_{n}}(x)dF_{\left|C_{ii}\right|^{2}}\left(x\right)\\
 & = & \int_{0}^{\infty}\lim_{n\rightarrow\infty}I_{\hat{\epsilon}_{n}}(x)dF_{\left|C_{ii}\right|^{2}}\left(x\right)\\
 & = & \int_{0}^{\infty}I_{0}(x)dF_{\left|C_{ii}\right|^{2}}\left(x\right)
\end{eqnarray*}
where $I_{0}(x)$ is the indicator function of the set $R_{0}=\{x\in\mathbb{R}:F_{\left|C_{ii}\right|^{2}}\left(x\right)-F_{\left|C_{ii}\right|^{2}}\left(\left(1-\epsilon\right)x\right)=0\}$.
Since we have assumed that $E\left[\left|C_{ii}\right|^{-2}\right]<\infty$,
also $\Pr(|C_{ii}|^{2}=0)=0$ and clearly $\mu_{F}(R_{0})=0$, and
$\mu(R_{\hat{\epsilon}_{n}})\rightarrow\mu(R_{0})$. 

\section{\label{sec:Proof-of-Lemma converge b}Proof of Lemma \ref{lem:converge b}}

To be explicit, let $X$ be a random variable on the probability space
$(\Omega,\mathcal{F},\mathbb{P})$ \cite{GrimmettBook}. Given the
definition of $X_{i}$, we can conclude that 

\begin{eqnarray*}
\lim_{i\to\infty}X_{i} & = & 0\qquad\mbox{w.p. 1.}
\end{eqnarray*}
Namely, if there is a set $B\in\mathcal{F}$ with $\mathbb{P}(B)>0$
where $\lim_{i\to\infty}X_{i}\neq0$ then $\mathbb{P}(B)\leq\mu_{X}\left(\bigcap_{i=1}^{\infty}G_{i}\right)$
which contradicts $\lim_{i\to\infty}\mu_{X}(G_{i})=0$.

Now $X_{i}\leq X$, and therefore by Lebesgue dominated convergence
\begin{eqnarray*}
\lim_{i\to\infty}E[X_{i}] & = & E[\lim_{i\to\infty}X_{i}]\;=\;0.
\end{eqnarray*}

\section{\label{sec:Proof-of-Property matching}Proof of Property \ref{lem:matching}}

Model the interference channel as a graph $G_{K}$, with $K=2M$ vertices
$u_{1},u_{2},\cdots,u_{2n}$. Vertices $u_{i}$ and $u_{j}$ are connected
by edge $E_{ij}$ if they form a weak$\left(1-\epsilon\right)$-pair,
i.e., $\left(1-\epsilon\right)\leq\frac{\left|C_{ij}\right|^{2}}{\left|C_{jj}\right|^{2}}\leq1\mbox{ or }\left(1-\epsilon\right)\leq\frac{\left|C_{ji}\right|^{2}}{\left|C_{ii}\right|^{2}}\leq1$.
Divide vertices into two disjoint classes $V_{1}=\left\{ u_{1},u_{2},\cdots,u_{M}\right\} $
and $V_{2}=\left\{ u_{M+1},u_{M+2},\cdots,u_{2M}\right\} $. Now define
event
\begin{eqnarray*}
\hat{A}_{\epsilon} & = & \left\{ \mathrm{there\; exists\; a\; perfect\; matching\; in\; the\; bipartite\; graph}\; G_{M,\, M}\right\} .
\end{eqnarray*}
As $\hat{A}_{\epsilon}\subseteq A_{\epsilon}$, $P\left(A_{\epsilon}\right)\geq P\left(\hat{A}_{\epsilon}\right)$.
Thus, if we can show that $P\left(\hat{A}_{\epsilon}\right)=1-o\left(1\right)$
as $K\rightarrow\infty$ then Property \ref{lem:matching} holds. 

For any bipartite graph, a perfect matching exists if Hall's condition
is satisfied.
\begin{thm}[Hall 1935]
\label{thm:Hall-1935} Given a bipartite graph \textup{$G_{M,\, M}$}
with disjoint vertices class $V_{1}$ and $V_{2}$, $V_{1}\bigcup V_{2}=V$,
$\left|V_{i}\right|=M$, whose set of edges is $E\left(G_{M,\, M}\right)$,
a perfect matching exists if and only if for every $S\subseteq V_{i},i=1or2$,
$\left|N\left(S\right)\right|\geq\left|S\right|$, where $N\left(S\right)=\left\{ y:\, xy\in E\left(G_{M,\, M}\right)\; for\; some\; x\in S\right\} $. 
\end{thm}
Any bipartite graph that does not have a perfect matching has following
properties
\begin{lem}
\label{lem:Lemma-7.12}Suppose $G_{M,\, M}$ has no isolated vertices
and it does not have a perfect matching. Then Hall's condition must
be violated by some set $A\subset V_{i}$, $i=1\, or\,2$. And such
set with minimal cardinality satisfies the following necessary conditions 

(i) $\left|N\left(A\right)\right|=\left|A\right|-1$;

(ii) $2\leq\left|A\right|\leq\left\lceil \frac{M}{2}\right\rceil $

(iii) the subgraph of G spanned by $A\bigcup N\left(A\right)$ is
connected, and it has at least $2a-2$ edges; 

(iv) every vertex in $N\left(A\right)$ is adjacent to at least two
vertices in $A$;

(v) any subsets of $N\left(A\right)$ can find a perfect matching
in $\left|A\right|$;\end{lem}
\begin{IEEEproof}
(i), (ii), (iii), and (iv) are proved by Lemma 7.12 in \cite{bollobas01},
and p.82 of \cite{Janson00randomgraphs}. And (iv) is true because
if there exists a subset $B$ of $N\left(A\right)$ that can not find
a perfect match, we could just let $B$ be $\hat{A}$, and its neighbors
in $A$ be $N\left(\hat{A}\right)$. Then $\hat{A}$ violates Hall's
condition, while $\left|\hat{A}\right|<a$. This contradicts the assumption
that $A$ is the minimal set violating Hall's condition.
\end{IEEEproof}
Define the event $F_{a}$: there is a set $A\subset V_{i}$, $i=1\, or\,2$,
$\left|A\right|=a$. satisfying (i), (ii) and (iii) in Lemma \ref{lem:Lemma-7.12}.
\cite{bollobas01} shows that for a graph with no isolated vertex,
$P\left(A_{\epsilon}\right)=1-o\left(1\right)$ is equivalent to $P\left(\bigcup_{a=2}^{\left\lceil \frac{M}{2}\right\rceil }F_{a}\right)=o\left(1\right)$.
Define $F_{1}$ as the event that there exists at least one isolated
vertex in $G_{M,M}$. In our case, we want to show that 
\begin{eqnarray*}
P\left(\bigcup_{a=2}^{\left\lceil \frac{M}{2}\right\rceil }F_{a}\right)+P\left(F_{1}\right) & = & o\left(1\right).
\end{eqnarray*}
Using the union bound, we have
\begin{eqnarray*}
P\left(F_{1}\right) & \leq & \sum_{i=1}^{2M}P\left(u_{i}\; isolated\right)\\
 & \leq & 2M\cdot P\left(u_{1}\; isolated\right)\\
 & \overset{\left(a\right)}{\leq} & 2M\cdot\left(1-p_{1j}\right)^{M}\\
 & \overset{\left(b\right)}{=} & o\left(1\right)
\end{eqnarray*}
where $p_{1j}\triangleq P\left(\left(1-\epsilon\right)\leq\frac{\left|C_{1j}\right|^{2}}{\left|C_{jj}\right|^{2}}\leq1\right)$,
$j=M+1,\cdots,2M$. We also define $p_{0}\triangleq P\left(\left(1-\epsilon\right)\leq\frac{\left|C_{ij}\right|^{2}}{\left|C_{jj}\right|^{2}}\leq1,or\left(1-\epsilon\right)\leq\frac{\left|C_{ji}\right|^{2}}{\left|C_{ii}\right|^{2}}\leq1\right)$
for later use. (a) holds because the event $\frac{\left|C_{1j}\right|^{2}}{\left|C_{jj}\right|^{2}}\notin\left(1-\epsilon,1\right)$
is independent of $j$. And it is a necessary condition for $V_{1}$
to be is isolated. 

Now, let us look into $F_{a}$ for $2\leq a\leq\left\lceil \frac{M}{2}\right\rceil $.
Let $A_{1}\subset V_{1}$, $A_{2}\subset V_{2}$, and $\left|A_{1}\right|=\left|A_{2}\right|+1=a$.
Denote $P\left(\mathcal{A}_{a}\right)$ as the probability that the
subgraph of $G_{M,M}$ spanned by $A_{1}\bigcup A_{2}$ satisfies
(i), (ii), and (iii) in Lemma \ref{lem:Lemma-7.12}. We have
\begin{eqnarray}
P\left(\bigcup_{a=2}^{\left\lceil \frac{M}{2}\right\rceil }F_{a}\right) & \overset{(d)}{\leq} & \sum_{a=2}^{\left\lceil \frac{M}{2}\right\rceil }P\left(F_{a}\right)\nonumber \\
 & \overset{(e)}{\leq} & 2\sum_{a=2}^{\left\lceil \frac{M}{2}\right\rceil }\left(\begin{array}{c}
M\\
a
\end{array}\right)\left(\begin{array}{c}
M\\
a-1
\end{array}\right)P\left(\mathcal{A}_{a}\right).\label{eq:pfa}
\end{eqnarray}
$(d)$ is from the union bound; $(e)$ is from the union bound, and
from that fact that there are $2\left(\begin{array}{c}
M\\
a
\end{array}\right)$ choices for $A$ with $\left|A\right|=a$, and $\left(\begin{array}{c}
M\\
a-1
\end{array}\right)$ more choices for $N\left(A\right)$. In \cite{bollobas01,Janson00randomgraphs},
the case where edge probabilities are i.i.d, whose value is $p$,
is considered. In \cite{bollobas01}, $P\left(\mathcal{A}_{a}\right)$
is bounded using condition (i), (ii) and (iii), which gives $P\left(\mathcal{A}_{a}\right)\leq\left(\begin{array}{c}
a\left(a-1\right)\\
2a-2
\end{array}\right)p^{2a-2}p^{a\left(n-a+1\right)}$. The term $p^{a\left(n-a+1\right)}$ is the probability that the
vertices in $A_{1}$ do not connect to vertices in $V_{2}-A_{2}$.
And in \cite{Janson00randomgraphs}, condition (iv) instead of (iii)
are used, which gives $P\left(\mathcal{A}_{a}\right)\leq\left(\begin{array}{c}
a\\
2
\end{array}\right)^{a-1}p^{2a-2}p^{a\left(n-a+1\right)}$. However, in our case, any two edges having adjacent vertices are
dependent. So we use condition (v). Since for $N\left(A\right)$,
a perfect match exists, then the subgraph spanned by $A\bigcup N\left(A\right)$
has $a-1$ edges that are not adjacent with each other. Thus, $P\left(\mathcal{A}_{a}\right)$
can be bounded by 
\begin{eqnarray}
P\left(\mathcal{A}_{a}\right) & \leq & Pr\left(\mathrm{condition\,(i),\,(ii)\, and\,(iv)\, are\, satisfied},\right.\label{eq:p a}\\
 &  & \left.\;\mathrm{vertices\, in\, A_{1}\, do\, not\, connect\, to\, vertices\, in}\; V_{2}-A_{2}\right)\\
 &  & \left(p_{0}^{a-1}\prod_{k=2}^{a}\left(\begin{array}{c}
k\\
1
\end{array}\right)\right)p_{A_{1}\bar{A}_{2}}
\end{eqnarray}
where 
\begin{eqnarray}
p_{A_{1}\bar{A}_{2}} & = & P\left(\frac{\left|C_{ij}\right|^{2}}{\left|C_{jj}\right|^{2}}\notin\left[\left(1-\epsilon\right),1\right],\;\mathrm{for\; all}\; u_{i}\in A_{1}\, and\, u_{j}\in\left(V_{2}-A_{2}\right)\right)\nonumber \\
 & \leq & \prod_{u_{j}\in\left(V_{2}-A_{2}\right)}P\left(\frac{\left|C_{ij}\right|^{2}}{\left|C_{jj}\right|^{2}}\notin\left[\left(1-\epsilon\right),1\right],\;\mathrm{for\; all}\; u_{i}\in A_{1}\right)\nonumber \\
 & = & \left(P\left(\frac{\left|C_{ji}\right|^{2}}{\left|C_{jj}\right|^{2}}\notin\left[\left(1-\epsilon\right),1\right],i=1,\cdots,\, a\; j=M+a\right)\right)^{M-a+1}\label{eq:paa}
\end{eqnarray}
notice that the event $\frac{\left|C_{ij}\right|^{2}}{\left|C_{jj}\right|^{2}}\notin\left[\left(1-\epsilon\right),1\right],\;\mathrm{for\; all}\; u_{i}\in A_{1}\, and\, u_{j}\in\left(V_{2}-A_{2}\right)$
is an necessary Substitute $\left|C_{j1}\right|^{2}$ by $x_{j}$,
and $\left|C_{11}\right|^{2}$ by $x_{1}$, denote their CDF by $F_{x}\left(x\right)$,
and their joint CDF $F_{\underbar{x}}\left(\underbar{x}\right)$.
Notice that $\left|C_{ij}\right|^{2}$ are i.i.d. distributed. Then
\begin{eqnarray}
 &  & P\left(\frac{\left|C_{ji}\right|^{2}}{\left|C_{jj}\right|^{2}}\notin\left[\left(1-\epsilon\right),1\right],i=1,\cdots,\, a\; j=M+a\right)\nonumber \\
 & = & \int_{A_{\underbar{x}}}dF_{\underbar{x}}\left(\underbar{x}\right)\nonumber \\
 & = & \int_{0}^{\infty}f_{x_{M+a}}\left(x_{M+a}\right)\prod_{i=1}^{a}\left(\int_{\frac{x_{i}}{x_{M+a}}\notin\left[\left(1-\epsilon\right),1\right]}f_{x_{i}}\left(x_{i}\right)dx_{i}\right)dx_{M+a}\nonumber \\
 & = & \int_{0}^{\infty}f_{x_{1}}\left(x_{1}\right)\left(\int_{\frac{x_{1}}{x_{M+a}}\notin\left[\left(1-\epsilon\right),1\right]}f_{x_{M+1}}\left(x_{M+1}\right)dx_{M+1}\right)^{a}dx_{M+a}\nonumber \\
 & \overset{(f)}{\leq} & \left(\int_{0}^{\infty}f_{x_{1}}^{2}\left(x_{1}\right)dx_{1}\right)^{\nicefrac{1}{2}}\left(\int_{0}^{\infty}g_{x_{1}}^{2}\left(x_{1}\right)dx_{1}\right)^{\nicefrac{a}{2}}\label{eq:p1j}
\end{eqnarray}
where $g\left(x_{1}\right)\triangleq\int_{\frac{x_{1}}{x_{M+a}}\notin\left[\left(1-\epsilon\right),1\right]}f_{x_{M+1}}\left(x_{M+1}\right)dx_{M+1}$.
(f) is from Cauchy-Schwartz inequality. Denote 
\begin{eqnarray*}
q_{1} & \triangleq & \left(\int_{0}^{\infty}f_{x_{1}}^{2}\left(x_{1}\right)dx_{1}\right)^{\frac{1}{2\left(M-a+1\right)}}\left(\int_{0}^{\infty}g_{x_{1}}^{2}\left(x_{1}\right)dx_{1}\right)^{\nicefrac{1}{2}}
\end{eqnarray*}
$q_{1}<1$, and $\lim_{M\rightarrow\infty}q_{1}=\left(\int_{0}^{\infty}g_{x_{1}}^{2}\left(x_{1}\right)dx_{1}\right)^{\nicefrac{1}{2}}$.
Notice that this limit value do not depend on the value of $M$. Now
combining (\ref{eq:paa}) and (\ref{eq:p1j}), we have
\begin{eqnarray}
P\left(\mathcal{A}_{a}\right) & \leq & \left(p_{0}^{a-1}\prod_{k=2}^{a}\left(\begin{array}{c}
k\\
1
\end{array}\right)\right)q_{1}^{a\left(M-a+1\right)}.\label{eq:paa 2}
\end{eqnarray}
Given (\ref{eq:pfa}) and (\ref{eq:paa 2}), 
\begin{eqnarray*}
P\left(\bigcup_{a=2}^{\left\lceil \frac{M}{2}\right\rceil }F_{a}\right) & \leq & 2\sum_{a=2}^{\left\lceil \frac{M}{2}\right\rceil }\left(\begin{array}{c}
M\\
a
\end{array}\right)\left(\begin{array}{c}
M\\
a-1
\end{array}\right)\left(p_{0}^{a-1}\prod_{k=2}^{a}\left(\begin{array}{c}
k\\
1
\end{array}\right)\right)q_{1}^{a\left(M-a+1\right)}\\
 & \leq & 2\sum_{a=2}^{\left\lceil \frac{M}{2}\right\rceil }\left(\frac{eM}{a}\right)^{a}\left(\frac{eM}{a-1}\right)^{a-1}a^{a-1}p_{0}^{a-1}q_{1}^{a\left(M-a+1\right)}\\
 & \leq & 2q_{1}\sum_{a=2}^{\left\lceil \frac{M}{2}\right\rceil }\left(\frac{e^{2}M^{2}}{\left(a-1\right)^{2}}ap_{0}q_{1}^{\frac{M}{2}}\right)^{a-1}\\
 & \leq & 2q_{1}\sum_{a=2}^{\left\lceil \frac{M}{2}\right\rceil }\left(2e^{2}p_{0}M^{2}q_{1}^{\frac{M}{2}}\right)^{a-1}\\
 & \leq & 2q_{1}\frac{M}{2}\left(2e^{2}p_{0}M^{2}q_{1}^{\frac{M}{2}}\right)\\
 & = & o\left(1\right).
\end{eqnarray*}
This means we can find a perfect matching with high probability, i.e.,
$1-o\left(1\right)$.
\end{document}